\documentclass[12pt]{article} %

\usepackage{subfig,varioref}

\IfFileExists{srcltx.sty}{\usepackage[active]{srcltx}}

\usepackage[sort&compress,comma,square,numbers]{natbib} %
\usepackage{amssymb,amsfonts,amsmath,amscd,float} \usepackage{graphicx}
\usepackage[a4paper,vscale=.8,hscale=.8,centering]{geometry}

\usepackage{xspace} 

\usepackage{hyperref}

\newcommand{\m}{s} 
\newcommand{\dir}{\textsc{d}} 
\newcommand{\numsm}{{$\nu$MSM}\xspace} 
\newcommand{\eV}{\:\mathrm{eV}} 
\newcommand{\ev}{\:\mathrm{eV}} 
\newcommand{\MeV}{\:\mathrm{MeV}} 
\newcommand{\mev}{\:\mathrm{MeV}} 
\newcommand{\GeV}{\:\mathrm{GeV}} 
\newcommand{\gev}{\:\mathrm{GeV}} 
 \newcommand{\diag}{\:\mathrm{diag}}
\newcommand{\im} {{\rm Im}} 
\newcommand{\p}{\partial} 

\makeatletter
\newcommand{\Rmnum}[1]{\expandafter\@slowromancap\romannumeral #1@}
\makeatother
\usepackage{setspace}

\begin{document}
\title{\rightline{\small \tt CERN-PH-TH-2011-312}Experimental bounds on
  sterile neutrino mixing angles}

\author{Oleg Ruchayskiy\thanks{CERN Physics Department, Theory Division,
    CH-1211 Geneva 23, Switzerland}~ and Artem
  Ivashko\thanks{Instituut-Lorentz for Theoretical Physics, Universiteit
    Leiden, Niels Bohrweg 2, Leiden, The Netherlands}
\thanks{Department of Physics, Kiev National Taras Shevchenko University, Glushkov str. 2 building 6, Kiev, 03022,
Ukraine}}

\date{}

\maketitle

\begin{abstract}
  We derive bounds on the mixing between the Standard Model (``active'')
  neutrinos and their right-chiral (``sterile'') counterparts in the see-saw
  models, by combining neutrino oscillation data and results of direct
  experimental searches.  We demonstrate that the mixing of sterile neutrinos
  with any active flavour can be significantly suppressed for the values of
  the angle $\theta_{13}$ measured recently by Daya Bay and RENO experiments.
  We reinterpret the results of searches for sterile neutrinos by the PS191
  and CHARM experiments, considering not only charged current but also neutral
  current-mediated decays, as applicable in the case of see-saw models.  The
  resulting \emph{lower bounds} on sterile neutrino lifetime are up to an
  order of magnitude \emph{stronger} than previously discussed in the
  literature.  %
  Combination of these results with the upper bound on the lifetime coming
  from primordial nucleosynthesis rule out the possibility that two sterile
  neutrinos with the masses between 10 MeV and the pion mass are solely
  responsible for neutrino flavour oscillations.
  We discuss the implications of
  our results for the Neutrino Minimal Standard Model (the $\nu$MSM).
\end{abstract}

\section{Introduction}
\label{sec:introduction}

Transitions between neutrinos of different flavours (see
e.g.~\cite{Strumia:06} for a review and
Refs.~\cite{Schwetz:08a,Schwetz:2011zk,Fogli:2011qn} for the recent update of
experimental values) are among the few firmly established phenomena
\emph{beyond the Standard Model of elementary particles}. The simplest
explanation is provided by the ``neutrino flavour oscillations'' --
non-diagonal matrix of neutrino propagation eigenstates in the weak charge
basis. While the absolute scale of neutrino masses is not established,
particle physics measurements put the sum of their masses below
$2\ev$~\cite{PDG:11} while from the cosmological data one can infer an upper
bound of $0.58\ev$ at 95\% CL ~\cite{WMAP7}.

A traditional explanation of the smallness of neutrino masses is provided by
the \emph{see-saw
  mechanism}~\cite{Minkowski:77,Ramond:79,Mohapatra:79,Yanagida:80}.  It
assumes the existence of several \emph{right-handed neutrinos}, coupled to
their Standard Model (SM) counterparts via the Yukawa interaction, providing
the Dirac masses, $M_\dir$, for neutrinos. The Yukawa interaction terms dictate
the SM charges of the right-handed particles: they turn out to carry no
electric, weak and strong charges; therefore they are often termed
``singlet,'' or ``sterile'' fermions.  Sterile neutrinos can thus have
Majorana masses, $M_\m$, consistent with the gauge symmetries of the Standard Model. If the
Majorana masses are much larger than the Dirac ones, the \emph{type I seesaw
  formula}~\cite{Minkowski:77,Ramond:79,Mohapatra:79,Yanagida:80} holds,
expressing the mass matrix of observed neutrinos $(\mathcal{M}_\nu)$ via
\begin{equation}
  \mathcal{M}_\nu = -
  M_\dir M_\m^{-1} M_\dir^T,
\label{see-saw}
\end{equation}
where $\mathcal{M}_\nu$ is a $3\times 3$ matrix of active neutrino masses,
mixings, and (possible) CP-violating phases. The masses of sterile neutrinos
are given by the eigenvalues of their Majorana mass matrix (with the
corrections of the order $M_\dir^2/M^2_\m$). They are much heavier than the
active neutrino masses as a consequence of~(\ref{see-saw}).

Numerous searches for sterile neutrinos in the mass range up to $\sim 100\gev$
had been performed (see the corresponding section in Particle Data
Group~\cite{PDG:11},\footnote{\url{http://pdglive.lbl.gov/Rsummary.brl?nodein=S077&inscript=Y}}
see also~\cite{Atre:09} and refs. therein).  These searches provided upper
bounds on the strength of interaction of these neutral leptons with the SM
neutrinos of different flavours -- active-sterile neutrino \emph{mixing
  angles} $\vartheta_\alpha^2 \propto
\left|\frac{M_{\dir,\alpha}}{M_\m}\right|^2$ for sterile neutrino with the
mass $M_\m$.\footnote{Here and below we use the letter $\vartheta$ for
  \emph{active-sterile mixing angles} (defined by Eq.~(\ref{eq:19}) below)
  while reserving $\theta_{12},\theta_{13}$ and $\theta_{23}$ for the measured
  parameters of the active neutrinos matrix $\mathcal{M}_\nu$.  These
  quantities $\vartheta_\alpha$ are often denoted $|V_{4\alpha}|^2$ or
  $|U_{x\alpha}|^2$ in the experimental papers, to which we refer. Here and
  below the Greek letters $\alpha,\beta$ are flavour index $e,\mu,\tau$ and
  $i,j=1,2,3$ denote active neutrino mass eigenstates.}  These bounds then can
be interpreted as \emph{lower bounds} on the lifetime of sterile neutrinos
$\tau_s$ via
\begin{equation}
  \label{eq:lifetime_expression}
  \tau_s^{-1} = \frac{G_F^2 M_\m^5}{96\pi^3}\sum_X  \vartheta_\alpha^2 B_{X}^{(\alpha)},
\end{equation}
where the sum runs over various particles to which sterile neutrino can decay,
depending on their mass ($\nu$, $e^\pm$,$\mu^\pm$,$\tau^\pm$, $\pi$, $K$,
heavier mesons and baryons) and dimensionless quantities $B_X^\alpha$ depend
{ on the branching ratios (see Appendix \ref{app:lifetime} for details).  The
lower bound on the lifetime $\tau_s$ is usually dominated by the least
constrained mixing angle, $\vartheta_\tau^2$ (as will be shown later).

This bound can be made \emph{stronger} if one assumes that the same particles
are also responsible for the neutrino oscillations. The see-saw
formula~(\ref{see-saw}) limits (at least partially) possible values of ratios
of the mixing angles $\vartheta_\alpha^2/\vartheta_\beta^2$. In the simplest
case when only two sterile neutrinos are present (the minimal number, required
to explain two observed neutrino mass differences) the ratios of mixing angles
varies within a limited range, see
e.g.~\cite{Gorbunov:07a,Shaposhnikov:08a}. While this range can be several
orders of magnitude large (owing to our ignorance of certain oscillation
parameters, such as e.g. a CP-violating
phase~\cite{Gorbunov:07a,Shaposhnikov:08a}), the implied (lower) bounds on the
lifetime become much stronger, essentially being determined by the
\emph{strongest}, rather than the weakest direct bound on $\vartheta_\alpha$.}
When confronted with the upper bound from Big Bang
Nucleosynthesis~\cite{Dolgov:00a,Dolgov:00b}, they seem to close the window of
parameters for sterile neutrinos with the mass lighter than about
$150\mev$~\cite{Gorbunov:07a,Boyarsky:09a}. It was argued
  in~\cite{Asaka:2011pb}, however, that in the case of normal hierarchy there
  can be a small allowed window of parameters of sterile neutrinos with the
  mass below the pion mass.

In this paper we reanalyze restrictions on sterile neutrino lifetime in view
of the recent results of the Daya Bay~\cite{An:2012eh} and
RENO~\cite{Ahn:2012nd} collaborations, that measured a non-zero mixing angle
$\theta_{13}$ (see also~\cite{Adamson:2011qu,Abe:2011sj}).  We demonstrate
that in the case when there are only two sterile neutrinos, responsible for
the observed neutrino oscillations, the oscillation data allow for such a
choice of the active-sterile Yukawa couplings that the mixing of sterile
neutrinos with any given flavour can be strongly suppressed. This happens
\emph{only} for a non-zero values of $\theta_{13}$, in the range consistent
with the current
measurements~\cite{Schwetz:08a,Fogli:2011qn,An:2012eh,Ahn:2012nd}.  We also
confront our results with the recently reanalyzed bounds from primordial
nucleosynthesis~\cite{Ruchayskiy:2012si} and show that the window in
the parameter space of sterile neutrinos with masses $10\MeV \lesssim M_s
  \lesssim 140\MeV$ discussed in previous works~\cite{Asaka:2011pb} (see
  also~\cite{Gorbunov:07a}) is closed. For larger masses the window remains
  open. The results of this paper partially overlap with~\cite{Asaka:2011pb}
(also~\cite{Gorkavenko:2009vd}), and we compare in the corresponding places to
the previous works.

The paper is organized as follows: in Section~\ref{sec:numsm} we briefly
describe the type I see-saw model we use. We then investigate the relations
between different mixing angles imposed by the see-saw mechanism and
demonstrate that the mixing with any flavour $\vartheta_\alpha^2$ can become
suppressed (Section~\ref{sec:solution-see-saw}). Section~\ref{sec:experiments}
is devoted to the overview of the experiments, searching for sterile neutrinos
with the masses below $2\GeV$, and the way one should interpret their results
to apply to the see-saw models that we study. Section~\ref{sec:discussion}
summarizes our revised bounds on mixing angles and translates them into the
resulting constraints on sterile neutrino lifetime
(Figs.~\ref{fig:lifetime_result}). We conclude in
Section~\ref{sec:conclusion}, discussing implications of our results and
confronting them with the bounds from primordial nucleosynthesis.

\section{Sterile neutrino Lagrangian}
\label{sec:numsm}

The minimal way to add sterile neutrinos to the Standard Model is provided by
the Type I see-saw
model~\cite{Minkowski:77,Ramond:79,Mohapatra:79,Yanagida:80} (see
also~\cite{Seesaw25,Schechter:1980gr,Schechter:1981cv,Rodejohann:2011vc} and refs. therein):
\begin{equation}
  \label{lagr1}
  \Delta\mathcal{L}_N = \sum_{I,J=1}^{\mathcal{N}} i \bar N_I \partial_\mu \gamma^\mu N_I -
  \left(F_{\alpha I} \,\bar L_\alpha N_I \tilde \Phi 
    - \frac{(M_\m)_{IJ}}{2} \; \bar {N_I^c} N_J + h.c.\right)~,
\end{equation}
where $F_{\alpha I}$ are new Yukawa couplings, $\Phi_i$ is the SM Higgs
doublet, $\tilde \Phi_i = \epsilon_{ij} \Phi^\dagger_j$. This model is
renormalizable, has the same gauge symmetries as the Standard Model, and
contains $\mathcal{N}$ additional Weyl fermions $N_I$ --- sterile neutrinos
($N_I^c$ being the charge-conjugate fermion, in the chiral representation of
Dirac $\gamma$-matrices $N_I^c = i\gamma^2 N_I^\dagger$).  The number of these
singlet fermions must be $\mathcal{N}\ge 2$ to explain the data on neutrino
oscillations.  In the case of $\mathcal{N}=2$ there are \emph{11 new
  parameters} in the Lagrangian~(\ref{lagr1}), while the neutrino mass/mixing
matrix $\mathcal{M}_\nu$ has 7 parameters in this case. The situation is even
more relaxed for ${\cal N}>2$.  The see-saw formula~(\ref{see-saw}) does not
allow to fix the scale of Majorana and Dirac $M_{\dir,\alpha I} = F_{\alpha I}
\langle\Phi\rangle$ masses.  In this work we will mostly concentrate on
sterile neutrinos with their masses $M_\m$ in the MeV--GeV range -- the range
that is probed by the direct searches. To further simplify our analysis we
will concentrate on the case when the masses of both sterile neutrinos are
close to each other (so that $\Delta M \ll M_\m$). One important example of
such model is provided by the \emph{Neutrino Extended Standard Model} (the
$\nu$MSM) (\cite{Asaka:05a,Asaka:05b}, see~\cite{Boyarsky:09a} for
review). The $\nu$MSM model contains 3 sterile neutrinos, whose masses are
roughly of the order of those of other leptons in the Standard Model.  Two of
these particles (approximately degenerate in their mass) are responsible for
baryogenesis and neutrino oscillations and the third one is playing the role
of dark matter.  The requirement of dark matter stability on cosmological
timescales makes its coupling with the Standard Model species so feeble, that
it does not contribute significantly to the neutrino oscillation
pattern~\cite{Boyarsky:06a}. Therefore, when analyzing neutrino oscillations,
the $N_1$ can be omitted from the Lagrangian and index $I$ in the sums runs
through 2 and 3 only:
\begin{equation}
\label{eq:nuMSM_Lagrangian}
\mathcal{L}_{see-saw} = \mathcal{L}_{SM} + \bar N_I i \p_\mu \gamma^\mu N_I -
M_{\dir,\alpha I} \bar \nu_\alpha N_I - M_{\dir,\alpha I}^* \bar N_I \nu_\alpha  - M_\m \bar N_2^c N_3 - M_\m \bar N_3 N^c_2 ~.
\end{equation}
This parametrization coincides with \cite[Eq.  (2.1)]{Shaposhnikov:08a}.  Note
that we use basis of singlet neutrinos where the mass matrix is off-diagonal.
Recent computation of the baryon asymmetry of the Universe in the
\numsm~\cite{Canetti:10a} demonstrated that sterile neutrinos with the mass as
low as several MeV can be responsible for baryogenesis and neutrino
oscillations within the $\nu$MSM. This prompts us to re-analyze the
implication of negative direct experimental searches for the Yukawa couplings
of sterile neutrinos with the MeV masses. We limit our analysis by $M_\m \le
2\gev$, as for the higher masses the existing experimental bounds do not probe
the region of mixing angles, required to produce successful baryogenesis in
the \numsm~\cite{Canetti:10a}.

\section{Solution of the see-saw equations }
\label{sec:solution-see-saw}
In this Section we investigate how mixing angles between active and sterile neutrinos are related to parameters of the observable neutrino matrix $\mathcal{M}_\nu$. We will demonstrate that the mixing angle $\vartheta_e^2$ in the case of normal
hierarchy and the mixing angles $\vartheta_\mu^2$ or $\vartheta_\tau^2$ in the
case of inverted hierarchy, can become suppressed as we vary the
parameters of the neutrino matrix away from their best-fit
values (but within the experimentally allowed $3\sigma$ bounds).

\subsection{Parametrization of the Dirac mass matrix}
\label{sec:param-solut-see}

We use the Pontecorvo--Maki--Nakagawa--Sakata (PMNS) parametrization of the
neutrino matrix $\mathcal{M}_\nu$ (see e.g. Eqs.(2.10) and (2.12) of
\cite{Strumia:06})
\begin{equation}
  \label{eq:22}
  \mathcal{M}_\nu = V^{*} \diag(m_1 e^{-2i\zeta},~m_2 e^{-2i\xi},~m_3) V^\dagger \;,
\end{equation}
where $V$ is the unitary matrix, whose explicit form is reminded in
Appendix~\ref{sec:pmns}.  Redefining a Dirac mass matrix as\footnote{The Dirac
  matrix $\tilde M_\dir$ has indexes $I=2,3$ and $i,j=1,2,3$ -- neutrino
  propagation basis}
\begin{equation}
  \label{eq:1}
  M_\dir \to \tilde M_\dir \equiv V^T M_\dir\;,
\end{equation}
we can rewrite the see-saw relation~(\ref{see-saw}) in the following form:
\begin{equation}
\label{eq:seesaw_relation_prime}
\diag(m_1 e^{-2i\zeta},~m_2 e^{-2i\xi},~m_3)_{ij} = -\frac{\tilde M_{\dir,i 2} \tilde M_{\dir,j 3} + \tilde M_{\dir,i 3} \tilde M_{\dir, j 2}}{M_\m}\;,
\end{equation}
The rank of the active neutrino mass matrix $\mathcal M_\nu$ is 2 in the case
of two sterile neutrinos, meaning that one of the masses $m_i$ is zero. Two
choices of ``hierarchies'' of the mass eigenvalues are possible. The first one
is called \emph{normal hierarchy} (NH) and corresponds to $0 \leq m_1 < m_2 <
m_3$. The second one is
called \emph{inverted hierarchy} (IH) and is realized for $0 \leq m_3 < m_1<
m_2$.

Once the mass $M_\m$ is fixed, the solutions of
Eq.~(\ref{eq:seesaw_relation_prime}) contain one unknown complex parameter,
$z$. Its presence reflects a symmetry of the see-saw relation
(\ref{eq:seesaw_relation_prime}) under the change $(\tilde M_{\dir,i 2},
\tilde M_{\dir,i 3}) \rightarrow (z \tilde M_{\dir, i 2}, z^{-1} \tilde
M_{\dir,i 3})$~\cite{Shaposhnikov:06b}.  It is this freedom that does not
allow to fix the absolute scale of $\tilde M_\dir$ (i.e. the value of
$\vartheta^2$) even if $M_\m$ is chosen.

The change $z \rightarrow z^{-1}$
is equivalent to the redefinition of $N_2 \rightarrow N_3,~N_3\rightarrow N_2$
together with shift of the Majorana phase $\xi\rightarrow \xi+\pi$
in~(\ref{eq:seesaw_relation_prime}).  Therefore in subsequent analysis we will
choose $|z|\ge 1$ without the loss of generality.

\begin{table}[t]
  \centering
  \begin{tabular}{ll|ll}
    \multicolumn{2}{c}{Normal hierarchy} & 
    \multicolumn{2}{c}{Inverted hierarchy}\\
    \hline
    {\Large\strut}$\Delta m^2_{21}$ & $ (7.09-8.19)\times 10^{-5} \eV^2$\\
    $\Delta m^2_{31}$ & $(2.14-2.76)\times 10^{-3} \eV^2$ & $\Delta m^2_{13}$ & $(2.13-2.67)\times 10^{-3} \eV^2$\\
    $\sin^2 \theta_{12} $ & $  0.27-0.36$\\
    $ \sin^2 \theta_{23} $ & $ 0.39-0.64$\\
    $\sin^2 \theta_{13} $ &  $0.010-0.038~(0.013-0.040)$\\ 
  \end{tabular}
  \caption{The $3\sigma$ bounds on the parameters of the mass matrix
    $\mathcal{M}_\nu$, adopted from~\protect\cite{Schwetz:2011zk,Fogli:2011qn,An:2012eh,Ahn:2012nd}. Here $\Delta
    m^2_{ij}=m_i^2-m_j^2$. The boundaries for inverted hierarchy are the same
    as for the normal one, unless written explicitly. The range of $\sin^2\theta_{13}$ is taken from the data of the Daya Bay experiment~\protect\cite{An:2012eh} (the values in parentheses -- from RENO~\protect\cite{Ahn:2012nd}).}
  \label{tab:3sigma_bounds}
\end{table}

\subsection{Normal hierarchy}
\label{sec:normal-hierarchy}

For normal hierarchy the explicit see-saw relation is
\begin{equation}
 \diag(0,~m_2 e^{-2i \xi},~m_3)_{ij} = -\frac{\tilde M_{\dir,i 2} \tilde M_{\dir,j 3} + \tilde M_{\dir,i 3} \tilde M_{\dir,j 2}}{M_\m}\;.
\end{equation}
Diagonal components of this matrix equation give
\begin{equation}
 \tilde M_{\dir,12} \tilde M_{\dir,13} = 0,~~\tilde M_{\dir,22} \tilde M_{\dir,23} = \frac{1}{2} m_2 M_\m e^{-2i\xi},~~\tilde M_{\dir,32} \tilde M_{\dir,33} = \frac{1}{2} m_3 M_\m\;.
\end{equation}
Using $m_2, m_3\neq 0$ we find that $\tilde M_{\dir,22},~\tilde
M_{\dir,23},~\tilde M_{\dir,32},~\tilde M_{\dir,33}$ are \emph{all} non-zero.
Analysis of non-diagonal terms reveals that \emph{both} $\tilde M_{\dir,12}$
and $\tilde M_{\dir,13}$ are zero and there are two general solutions
(c.f.~\cite{Shaposhnikov:06b}):
\begin{equation}
\label{eq:seesaw_solution_normal}
\tilde M^{\pm}_{\dir,i 2}=iz \sqrt{\frac{M_\m}{2}}(0,~\pm ie^{-i\xi}
\sqrt{m_2},~\sqrt{m_3}),~~\tilde M^{\pm}_{\dir,i 3}=i z^{-1}\sqrt{\frac{M_\m}{2}}(0,~\mp ie^{-i\xi} \sqrt{m_2},~\sqrt{m_3})\;.
\end{equation}
The solution $\tilde M^{+}_{\dir}$ with $\xi = \psi + \pi$ equals to $\tilde
M^{ -}_\dir$ with $\xi=\psi$. It allows us to consider only one solution
$\tilde M^{ +}_D$ on the interval $0\leq \xi<2\pi$. In what follows we
therefore omit the superscript~$+$.\footnote{Unlike the parametrizations used
  e.g.  in Ref.~\cite{Casas:01,Shaposhnikov:06b,Asaka:2011pb} this way of
  parametrizing the solution of the see-saw equations shows that there is only
  one branch of solutions, with all other related to it via redefinitions
  $N_2\leftrightarrow N_3$ and shift of the Majorana phases. In particular in
  the parametrization we used it is much easier to analyze whether mixing
  angles become zero. The relation $|z|=\exp(\im\: \omega)$ holds, where the
  parameter $\omega$ was employed in \cite{Asaka:2011pb}.}

The  mixing angles of the active-sterile neutrinos are defined as follows:
\begin{equation}
\label{eq:19}
2\vartheta_\alpha^2 \equiv\sum \limits_I |(M_\dir  M_\m^{-1})_{\alpha I}|^2 =\sum \limits_I |( V^*  \tilde M_\dir  M_\m^{-1})_{\alpha I}|^2 = \frac{1}{M_\m^2} \sum \limits_I |( V^* \tilde M_\dir)_{\alpha I}|^2\;.
\end{equation} 
Inserting the explicit solution~(\ref{eq:seesaw_solution_normal}) for $\tilde
M_\dir$ results in
\begin{equation}
  \vartheta_\alpha^2 = \frac{|z|^2}{4 M_\m} \left| \sqrt{m_3} V_{\alpha3} - i
    e^{i\xi} \sqrt{m_2} V_{\alpha 2} \right|^2  + \frac1{4 M_\m|z|^2} \left| \sqrt{m_3} V_{\alpha3} + i
    e^{i\xi} \sqrt{m_2} V_{\alpha 2} \right|^2.
\label{eq:theta_alpha-NH}
\end{equation}
For $|z|\gg1$ the contribution of $\tilde M_{\dir,i 3}$ is suppressed
compared with that of $\tilde M_{\dir,i 2}$ and therefore we neglect the former
(we will comment below on the case $|z| \gtrsim 1$).

As the value of the Majorana phase $\xi$ is undetermined experimentally, the
condition $\vartheta_\alpha=0$ is satisfied iff $m_3 |V_{\alpha3}|^2 = m_2
|V_{\alpha2}|^2$ (we neglect second term on the r.h.s. of (\ref{eq:theta_alpha-NH})). For the electron flavour ($\alpha=e$) it translates into
\begin{equation}
\label{eq:thetae_0_normal}
\sin^2 \theta_{12} \frac{m_2}{m_3} = \tan^2 \theta_{13}, 
\end{equation}
which, in principle, can be satisfied only for non-zero $\theta_{13}$. This
result has been already obtained in \cite{Asaka:2011pb}.

The bounds on the parameters of the mass matrix $\mathcal{M}_\nu$ at the
$3\sigma$ level that we use are shown in Table~\ref{tab:3sigma_bounds}. Note
that in this paper we do not take into account statistical correlations
between different oscillation parameters, allowing them to vary independently
within their $3\sigma$ intervals. Consequently, we obtain the $3\sigma$
intervals for the combinations of parameters, entering
Eq.~(\ref{eq:thetae_0_normal})\footnote{\label{fn:Daya-RENO-convention}
  Throughout this paper whenever two numbers are given instead of one, the first is based on the
  results of the Daya Bay experiment~\cite{An:2012eh}, and the second one (in
  parentheses) is obtained based on the result of application of the RENO
  bounds~\cite{Ahn:2012nd} (see Table~\ref{tab:3sigma_bounds}).}:
\begin{equation}
  \begin{array}{rcccl}
    0.043 &<& \sin^2 \theta_{12} \frac{m_2}{m_3} &<& 0.070,\\
  0.010 (0.014) &<& \tan^2\theta_{13} &<& 0.039 (0.042),
  \end{array}
\end{equation}

They imply that the relation (\ref{eq:thetae_0_normal}) {\it does not} hold exactly for
the neutrino oscillation parameters, presented in
Table~\ref{tab:3sigma_bounds}. Therefore the mixing angle $\vartheta_e^2$
cannot become zero, but has a non-trivial lower bound.  To find the minimal
value that it can reach, we consider the ratio of the angles
$\vartheta_e^2/(\vartheta_e^2+\vartheta_\mu^2+\vartheta_\tau^2)$. Due to the
unitarity of~$V$, the denominator is
\begin{equation}
\label{eq:sum_angles_normal}
\sum\limits_\alpha \vartheta_\alpha^2 \approx \frac{1}{2 M_\m^2} \sum\limits_{\alpha,\beta,\gamma}  V^*_{\alpha \beta} \tilde M_{\dir,\beta2} V_{\alpha \gamma} \tilde M^{*}_{\dir,\gamma2} = \frac{1}{2M_\m^2} \sum\limits_\beta \tilde |M_{\dir,\beta2}|^2 =\frac{|z|^2}{4 M_\m}(m_2+m_3). 
\end{equation}
Let us denote the ratio of the mixing of sterile neutrinos with one flavour to
the sum of all mixings by $T_\alpha$, 
\begin{equation}
  \label{eq:17}
  T_\alpha \equiv \frac{\vartheta_\alpha^2}{\sum\limits_\beta \vartheta_\beta^2}.
\end{equation}
Then we get the following expression for $T_e$:
\begin{equation}
\label{eq:bounds-theta_e-o-sum-NH}
T_e = \frac{ |i
  e^{i\xi} c_{13} s_{12} \sqrt{\frac{m_2}{m_3}} -
  s_{13}|^2}{1+\frac{m_2}{m_3}}. 
\end{equation}
The minimum is achieved if we push $\theta_{12}$ and $\Delta m_{21}^2$ to
their $3\sigma$ lower boundaries, $\theta_{13}$ and $\Delta m^2_{31}$ to their
upper boundaries, and choose $\xi=-\pi/2$. The maximum is achieved when we set
$\Delta m^2_{31}$ equal to its lower bound, $\theta_{13},\theta_{12}$ and
$\Delta m^2_{21}$  to their upper bounds, and by choosing the Majorana phase $\alpha =\pi/2$.
 The bounds on
$T_e$ from Table \ref{tab:mixing-ratio-bounds} translate into the bound for the muon and tau flavours combined:
\begin{equation}
  0.83 \leq T_\mu + T_\tau.
\end{equation}

The minimum and maximum of different $T_\alpha$ are listed in the
Table~\ref{tab:mixing-ratio-bounds} and in Fig.~\ref{fig:Talpha-min}.

This analysis was conducted in approximation of large $|z|$. See
Appendix~\ref{sec:z-1corrections} for the account of finite-$|z|$ effects.

\begin{table}
  \begin{tabular}[c]{|r|r|}
    \hline
    Normal hierarchy & Inverted hierarchy\\
    \hline
    $T_e \leq 0.15$ & $0.02 \leq T_e \leq 0.98$ \\
    $0.09\leq T_\mu \leq 0.89$ & $0 \leq T_\mu \leq 0.60$ \\
    $0.08\leq T_\tau \leq 0.88$ & $2\times 10^{-4}~(7\times 10^{-5}) \leq T_\tau \leq 0.62$ \\
    \hline
    \multicolumn{2}{|c|}{\sl The ranges are based on $2\sigma$ bounds}\\
    \hline
  \end{tabular} 
  \begin{tabular}[c]{|r|r|}
    \hline
    Normal hierarchy & Inverted hierarchy\\
    \hline
    $T_e \leq 0.17$ & $0.02 \leq T_e \leq 0.98$ \\
    $0.07 \leq T_\mu \leq 0.92$ & $0\leq T_\mu \leq 0.63$\\
    $0.06 \leq T_\tau \leq 0.90$ & $0\leq T_\tau \leq 0.65$\\
    \hline
    \multicolumn{2}{|c|}{\sl The ranges are based on $3\sigma$ bounds}  \\
    \hline
  \end{tabular}
  \caption{The ratio of the sterile neutrino mixing with a given flavour $\alpha$
    to the
    \emph{sum} of the three mixings, $T_\alpha$ (defined
    by~(\protect\ref{eq:17})).  \textbf{Left} table shows the upper and lower values of $T_\alpha$ when
    parameters of neutrino oscillations are allowed to vary within their $2\sigma$
    boundaries (taken from \protect\cite{Schwetz:2011zk}). The \textbf{right} table shows the results when the  parameters
    of active neutrino oscillations are varied within their $3\sigma$ limits
    (see Table~\protect\ref{tab:3sigma_bounds}). For the explanation of
      the numbers in parentheses, see Footnote~\protect\ref{fn:Daya-RENO-convention}. }
\label{tab:mixing-ratio-bounds}
\end{table}

\subsection{Inverted hierarchy}
\label{sec:inverted-hierarchy}

Similarly to the previous case, for the inverted hierarchy we get a solution
of the see-saw equations~(\ref{eq:seesaw_relation_prime})
\begin{equation}
\label{eq:seesaw_solution_inverted}
\tilde M_{\dir,i 2}=iz \sqrt{\frac{M_\m}{2}}(e^{-i\zeta}\sqrt{m_1},~ie^{-i\xi} \sqrt{m_2},~0),~~\tilde M_{\dir,i 3}=iz^{-1} \sqrt{\frac{M_\m}{2}}(e^{-i\zeta}\sqrt{m_1},~-ie^{-i\xi} \sqrt{m_2},~0)
\end{equation}
for $0\leq \xi < 2\pi$.  In this case $\vartheta_\mu^2$ or $\vartheta_\tau^2$
can become very suppressed, as we will show soon.

 The mixing angles are
\begin{equation}
\label{eq:mixing_angles_inverted}
\vartheta_\alpha^2 = \frac{|z|^2}{4 M_\m} \left| \sqrt{m_1}V_{\alpha1} - i e^{i(\xi-\zeta)} \sqrt{m_2} V_{\alpha2} \right|^2 + \frac{1}{4 M_\m|z|^2} \left| \sqrt{m_1}V_{\alpha1} + i e^{i(\xi-\zeta)} \sqrt{m_2} V_{\alpha2} \right|^2.
\end{equation}
For $|z|\gg 1$ they can become close to zero only if $\sqrt{m_1}
|V_{\alpha1}|= \sqrt{m_2} |V_{\alpha2}|$. For $\alpha=\mu$ this condition
translates into
\begin{equation}
\label{eq:thetamu_0_inverted}
 |\tan \theta_{12} + \sin\theta_{13}\tan\theta_{23} e^{-i\phi}| = \sqrt{\frac{m_2}{m_1}} |1- \sin\theta_{13}\tan\theta_{12} \tan\theta_{23} e^{-i\phi}|.
\end{equation}
For the parameter set close to the best fit, left-hand side is \textit{less}
than the right-hand side, because then $\sin \theta_{13} \approx 0$, while
$\tan \theta_{12}< 1$ and $m_1 \approx m_2$. To attain the equality one has to
push left-hand side up and the right-hand side down. $\phi=0$ makes phases of
both complex terms inside $|...|$ on the left-hand side equal, thereby the
absolute value of their sum becomes maximal. Simultaneously the right-hand
side becomes minimal. For this specific choice of the Dirac angle the equality
(\ref{eq:thetamu_0_inverted}) turns into
\begin{equation}
\label{eq:theta_mu-suppression-IH}
 \frac{\sqrt{\frac{m_2}{m_1}}-\tan\theta_{12}}{\sqrt{\frac{m_2}{m_1}}\tan\theta_{12}+1} = \sin\theta_{13}\tan\theta_{23}.
\end{equation}
The $3\sigma$ bounds for inverted hierarchy in general are the same as for the
normal one (see Table~\ref{tab:3sigma_bounds}) with the exception of the
``atmospheric'' mass difference, that slightly differs.  Using these values
we find
\begin{equation}
  \label{eq:2}
  0.14< \frac{\sqrt{\frac{m_2}{m_1}}-\tan\theta_{12}}{\sqrt{\frac{m_2}{m_1}}\tan\theta_{12}+1} < 0.24,\quad 0.08~(0.09)<\sin\theta_{13}\tan\theta_{23}<0.26.
\end{equation}
We see that two regions overlap, therefore the relation (\ref{eq:thetamu_0_inverted}) can be satisfied and
$\vartheta_\mu^2$ can be zero in a wide region of values of the parameters of
the neutrino oscillation matrix. See, however, Sec. \ref{sec:minimal-mixing} below.

Similarly, the condition $\vartheta_\tau=0$ (for $\phi=\pi$) translates into
\begin{equation}
\label{eq:3}
  \frac{\sqrt{\frac{m_2}{m_1}}-\tan\theta_{12}}{\sqrt{\frac{m_2}{m_1}}\tan\theta_{12}+1} = \sin\theta_{13}\cot\theta_{23},
\end{equation}
and can be satisfied, because the quantity on the right hand side varies from
$0.07~(0.09)$ to $0.24~(0.25)$, well within the range of~(\ref{eq:2}).\footnote{It was
  pointed out in~\cite{Asaka:2011pb} that both $\vartheta_\mu$ and
  $\vartheta_\tau$ can be suppress in inverted hierarchy, for
  $\theta_{13}=0$. For this to happen the relation
  $\sqrt{\frac{m_2}{m_1}}=\tan\theta_{12}$ should hold, as one can also see
  from Eqs.~(\ref{eq:theta_mu-suppression-IH}) and (\ref{eq:3}). The
  corresponding value of $\theta_{12}$ is however well outside the $3\sigma$
  interval. The general case $\theta_{13}\neq 0$ has not been analyzed
  in~\cite{Asaka:2011pb}.}

On the other hand, $\vartheta_e$ can be zero only if
\begin{equation}
  \label{eq:5}
  \cot \theta_{12} = \sqrt{\frac{m_2}{m_1}}
\end{equation}
can be realized. The left hand side is always larger than the right hand side (within the
$3\sigma$ region), therefore no $\vartheta_e$ suppression can occur. However
it is important to know what minimal value this mixing angle can
reach. According to Eq.(\ref{eq:mixing_angles_inverted}) electron mixing angle
is given by
\begin{equation}
  \vartheta_e^2 = \frac{|z|^2}{4 M_\m} \cos^2\theta_{13} \left( m_1 \cos^2\theta_{12} + m_2 \sin^2\theta_{12} + \sin(\xi-\zeta) \sin 2\theta_{12} \sqrt{m_1 m_2} \right) .
\end{equation}
For $\xi-\zeta=-\pi/2$ this quantity is minimal
\begin{equation}
 \vartheta^2_{e,min} = \frac{|z|^2}{4 M_\m^2} \cos^2\theta_{13} \left( \sqrt{m_1} \cos\theta_{12} - \sqrt{m_2} \sin\theta_{12} \right)^2.
\end{equation}
To compare it with the other mixing angles, we note that the relation
\begin{equation}
\label{eq:sum_angles_inverted}
\sum\limits_\alpha \vartheta_\alpha^2 \approx \frac{|z|^2}{4 M_\m}(m_1+m_2)
\end{equation}
holds (similar to Eq.(\ref{eq:sum_angles_normal}) in the case of normal
hierarchy).  Therefore
\begin{equation}
 \frac{\vartheta_{e,min}^2}{\sum\limits_\alpha \vartheta_\alpha^2} =
 \frac{\cos^2\theta_{13}}{1+\frac{m_2}{m_1}} \left( \cos\theta_{12} -
   \sqrt{\frac{m_2}{m_1}} \sin\theta_{12} \right)^2.
\end{equation}

\begin{figure}
\centering
\includegraphics[width=.55\textwidth]{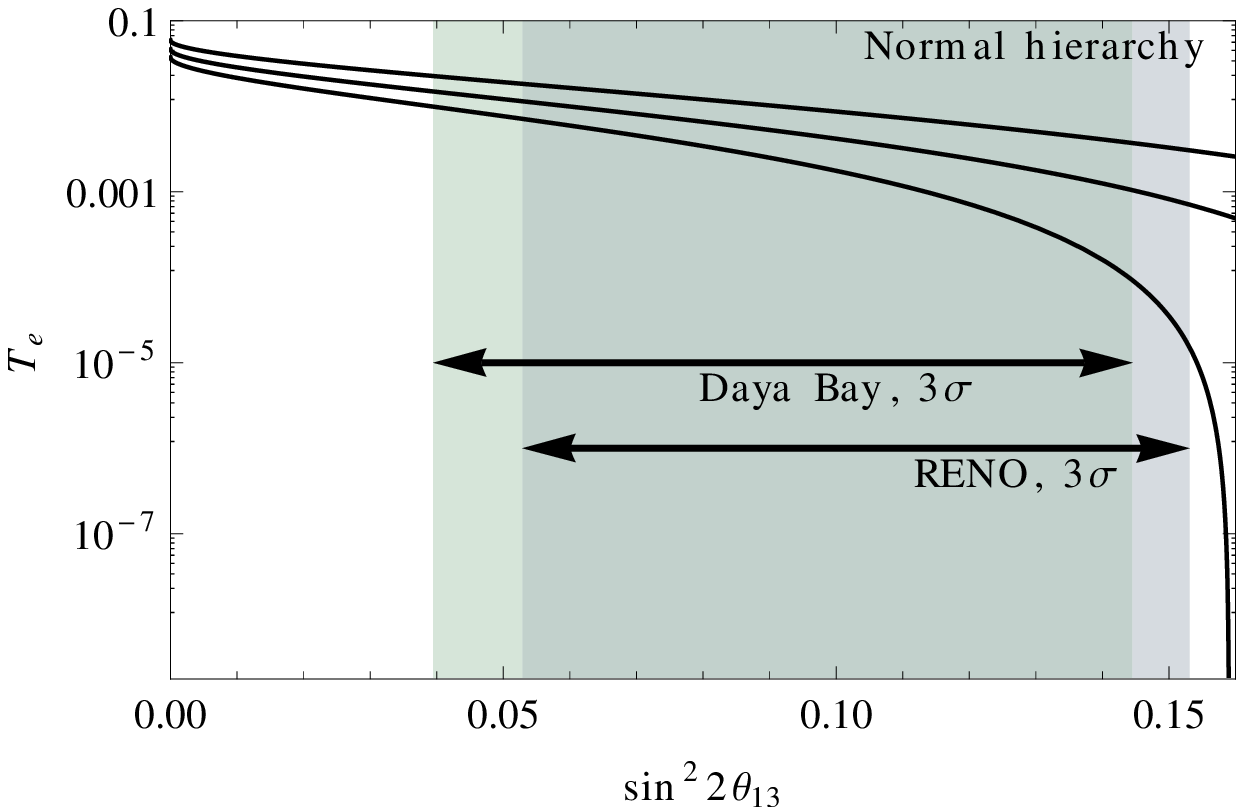}
\includegraphics[width=.49\textwidth]{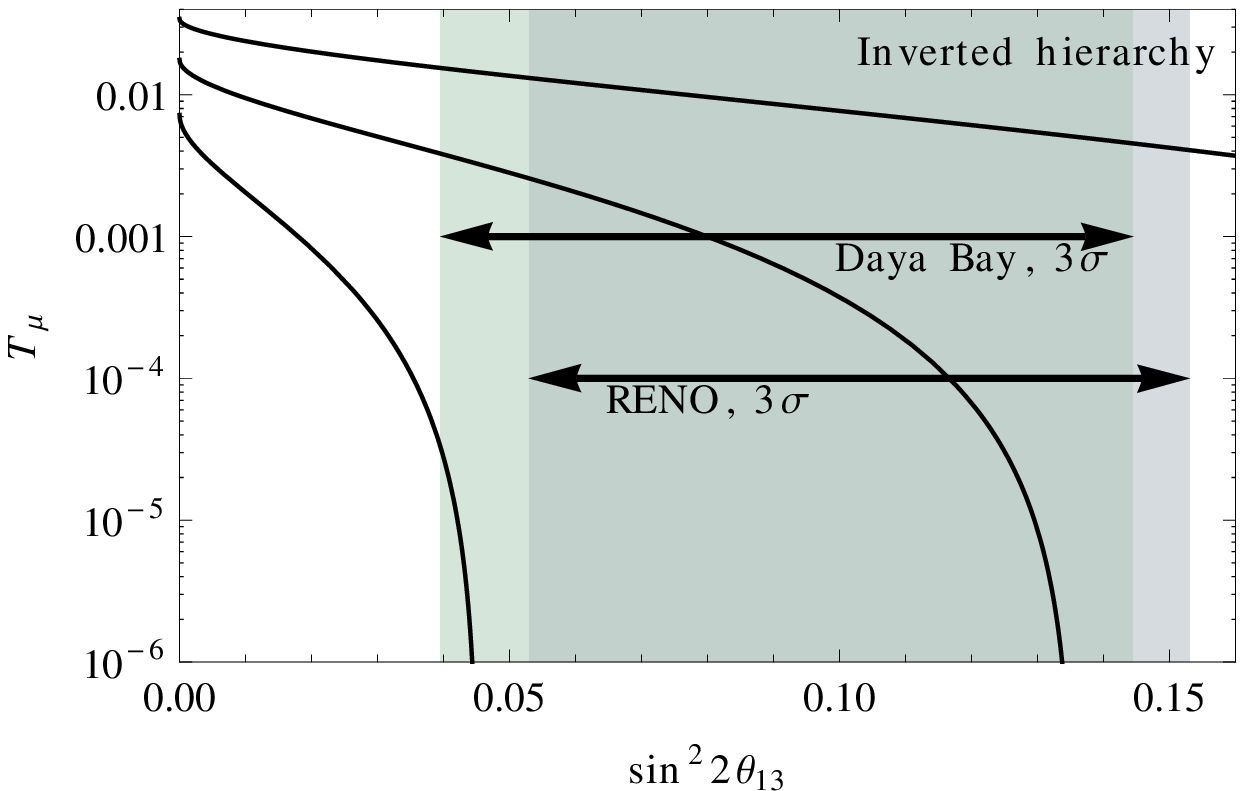}~
\includegraphics[width=.49\textwidth]{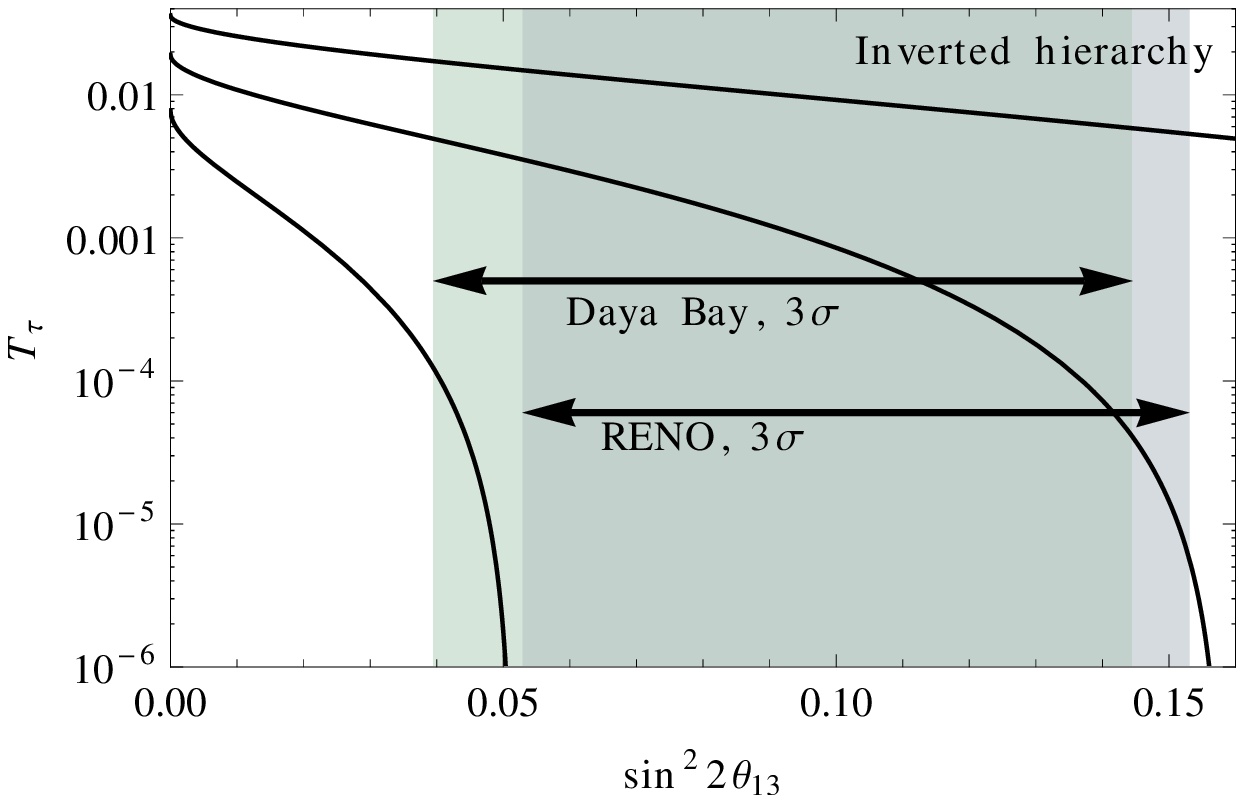}
\caption{The minimal ratios of mixing angles $T_\alpha = \vartheta_\alpha^2/\sum \vartheta_\beta^2$. \textbf{The upper} figure depicts normal hierarchy, \textbf{two lower} ones -- IH. In all figures, the lower curve corresponds to the choice of the mixing angles and mass splittings that minimizes the ratio within the $3\sigma$ range, upper -- that maximizes it, middle employs the best-fit parameters (for details of the choices, see Secs.~\protect\ref{sec:normal-hierarchy},~\protect\ref{sec:inverted-hierarchy}). CP-phases are $\xi=-\pi/2$ for the $T_e$-plot, $\phi=0,~\xi-\zeta=\pi/2$ for $T_\mu$, and $\phi=\pi,~\xi-\zeta=-\pi/2$ for $T_\tau$. The bands \emph{Daya Bay} and \emph{RENO} correspond to the $3\sigma$ ranges of $\theta_{13}$, indicated by the corresponding experiments~\protect\cite{An:2012eh,Ahn:2012nd}. }
\label{fig:Talpha-min}
\end{figure}

The results of the analysis are listed in Table \ref{tab:mixing-ratio-bounds} and Fig.~\ref{fig:Talpha-min}.
From the upper bound on $T_\alpha$ we derive the bound
\begin{equation}
\label{eq:4}
T_\mu+T_\tau \geq 0.02\;.
\end{equation}
We see that in this mass hierarchy it is possible for the overall coupling of the sterile neutrino to both $\mu$ and
$\tau$ flavours to become tiny compared to the electron flavour coupling.

\subsection{Minimal mixing angles in the \numsm}
\label{sec:minimal-mixing}

Finally, we find the \emph{minimal} values of the sterile neutrino mixing angles in
the \numsm, compatible with the neutrino oscillation data. These angles will
turn out to be much smaller than the experimental upper bounds in all regions
of masses, probed by the experiments. A general
solution of the see-saw equations
(\ref{eq:theta_alpha-NH}),~(\ref{eq:mixing_angles_inverted}) gives $\vartheta$
as a function of $|z|$:
\begin{equation}
\label{eq:11}
\vartheta^2_\alpha = A_\alpha |z|^2+ \frac{B_\alpha}{|z|^2}
\end{equation}
with coefficients $A_\alpha$ and $B_\alpha$ independent of $|z|$. The minimum
of this expression is reached for $|z|^2_\alpha=\sqrt{B_\alpha/A_\alpha}\ge 1$
and is given by
\begin{equation}
  \label{eq:6}
  (\vartheta^2_\alpha)_\text{min} = 2 \sqrt{A_\alpha B_\alpha}.
\end{equation}
To find an {\it absolute} lower bound on the mixing angle for a given sterile
neutrino mass, we vary this expression over the parameters of neutrino
oscillations. The resulting mixing angles and the corresponding values of
$|z|$ are listed in Table~\ref{tab:minimal}.\footnote{Notice, that the ratio
  of the mixing angles $\vartheta_\alpha^2/\vartheta_\beta^2$ does not reach
  its minimum when~(\ref{eq:6}) is satisfied. The values of $|z|$ for which
  the bounds on the lifetime are relaxed the most are those when some of the
  mixing angles reach their upper experimentally allowed value. } One
  can see that the values presented therein do not depend significantly on the
  $3\sigma$ upper bound on $\theta_{13}$ that we choose. The only exception is
  the minimum of the $\vartheta_e$ angle. In this case the exact value of the
  upper bound on $\theta_{13}$ defines how close $A_\alpha$, and hence
  $(\vartheta_\alpha^2)_\text{min}$, can come to zero.

\begin{table}[t]
  \centering
  \subfloat[NH, best-fit][NH, best-fit]{
 \begin{tabular}[c]{|c||c|c|}
   \hline
   Flavour $\alpha$ & $(\vartheta^2_\alpha)_\text{min}$ @ $1\mev$&  $|z|$   \\
   \hline
   $e$ & $7\times 10^{-10}$ & $2.2~(2.4)$ \\
   $\mu,\tau$ & $10^{-8}$ & $1.5$  \\
   \hline
    \end{tabular}
  }
  \qquad
  \subfloat[NH, $3\sigma$][NH, $3\sigma$]{
    \begin{tabular}[c]{|c||c|c|}
      \hline
      Flavour $\alpha$ & $(\vartheta^2_\alpha)_\text{min}$ @ $1\mev$ & $|z|$   \\
      \hline
      $e$ & $10^{-10}~(4\times 10^{-11})$ & $6.2~(9.8)$ \\
      $\mu$ & $8\times 10^{-10}~(6\times 10^{-10})$ & $5.4~(6.3)$ \\
      $\tau$ & $1.2\times 10^{-9}~(1.0\times 10^{-10})$ & $4.6~(5.1)$ \\
      \hline
    \end{tabular}
  }
  \\
  \subfloat[IH, best-fit][IH, best-fit]{
    \begin{tabular}[c]{|c||c|c|}
      \hline
      Flavour $\alpha$ & $(\vartheta^2_\alpha)_\text{min}$ @ $1\mev$ & \strut $|z|$   \\
      \hline
      $e$ & $10^{-8}$ & $2.3$ \\
      $\mu,\tau$ & $2\times 10^{-9}$ & $3.5$  \\
      \hline
    \end{tabular}
  }
  \qquad
  \subfloat[IH, $3\sigma$][IH, $3\sigma$]{
    \begin{tabular}[c]{|c||c|c|}
      \hline
      Flavour $\alpha$ & $(\vartheta^2_\alpha)_\text{min}$ @ $1\mev$ & \strut $|z|$   \\
      \hline
      $e$ & $6\times 10^{-9}$ & $2.7$ \\
      $\mu,\tau$ & \multicolumn{2}{c|}{see text}  \\
      \hline
    \end{tabular}
}
\caption{Minimal values of the active-sterile mixing angles
  $\vartheta_\alpha^2$, obtained using the best-fit 
  values of neutrino oscillation parameters or by
  varying the neutrino oscillation data within their $3\sigma$ intervals,
  listed in Table~\protect\ref{tab:3sigma_bounds}. The values for $(\vartheta^2_\alpha)_\text{min}$
  are provided for 
  sterile neutrinos with the mass $M_s=1\MeV$. For other masses one should multiply them by $(\MeV / M_s)$. Columns ``$|z|$''
  show the values of $|z|$ for which the minimum in~(\protect\ref{eq:11}) is reached. For the explanation of numbers in brackets, see Footnote~\protect\ref{fn:Daya-RENO-convention}.} 
 \label{tab:minimal}
\end{table}

For the mixing angles $\vartheta^2_{\mu,\tau}$ in the case of inverted
hierarchy $A_\alpha = 0, B_\alpha\neq 0$ and formally for infinitely large
$|z|$ they would become zero. The value of $|z|$, however, is bounded from
above, $|z| < z_\text{max}$, by the requirement that \emph{none of three}
mixing angles exceeds its upper bound (for quantitative estimates of
$z_\text{max}$, look at Fig.~\ref{fig:epsilon_bounds}). Therefore the couplings to $\mu$ and $\tau$ neutrinos remain
  \emph{finite}. Estimates of mixing
angles can be provided for $B_\alpha$ given by $L^{IH}_\alpha$
(\ref{eq:L_alpha-IH},\ref{eq:L_alpha-bounds}), along with $A_{\mu,\tau}=0$,
$z=z_\text{max}$
\begin{equation}
  \vartheta_{\mu,\tau}^2 \gtrsim 2\times 10^{-8} \frac{\MeV}{M_\m
    z^2_\text{max}},\qquad(\text{IH}) 
\label{eq:min-mutau-angle-estimate-IH}
\end{equation}

\section{Experimental bounds on sterile neutrino mixings}
\label{sec:experiments}

The direct experimental searches for neutral leptons had been performed by a
number of collaborations~\cite{Bernardi:1985ny,Bernardi:1987ek,%
  Britton:1992pg,Britton:1992xv,Vaitaitis:1999wq,CHARM:1985,Yamazaki:1984sj,Hayano:1982wu,Bryman:1996xd,Abela:1981nf,Daum:1987bg,Aoki:2011vma,Abreu:1996pa}
(see e.g.~\cite{Gorbunov:07a,Atre:09} for review of various constraints).  The
negative results of the searches are converted into the \emph{upper} bound on
$\vartheta_\alpha \vartheta_\beta$ for different flavours. If neutrino
oscillations are mediated by these sterile neutrinos, these bounds can be
translated into the \emph{upper} bounds on parameter $|z|$ and \emph{lower}
bounds on sterile neutrino lifetime.

Below, we take a closer look at two main types of experiments (``peak
searches'' and ``fixed target experiments'')\footnote{The neutrinoless
  double-beta decay ($0\nu\beta\beta$) does not provide significant
  restrictions on the parameters of the sterile neutrinos in the type-I
  see-saw models (contrary to the case discussed in e.g.~\cite{Atre:09}), see
  discussion in~\cite{Asaka:2011pb,Blennow:2010th}. In particular, this is the
  case in the \numsm~\cite{Bezrukov:2005mx}.}  and describe
\emph{reinterpretation of these bounds} in the case, when sterile neutrinos
with MeV--GeV masses are also responsible for neutrino oscillations.

\subsection{Peak searches}
\label{sec:peak-searches}

In ``\emph{peak search}'' experiments~\cite{Shrock:80,Shrock:1980ct,Shrock:1981wq,Shrock:1981cq}, one considers the
two-body decay of charged $\pi$ or $K$ mesons to charged lepton ($e^\pm$ or
$\mu^\pm$) and neutrino (see e.g.~\cite{Atre:09} for discussion).
In case of the pion decay the limit on $\vartheta_e^2$ for masses in the range
$60\mev \le M_\m \le 130\mev$ is provided by the searches for the secondary
positron peak in the decay $\pi^+ \to e^+ N$ to the massive sterile neutrino
$N$ as compared to the primary peak coming from the $\pi^+ \to e^+ \nu_e$
decay.  Recent analysis of~\cite{Aoki:2011vma} puts this limit at
$\vartheta_e^2 < 10^{-8}$ in the mass range $60-129\mev$, for earlier results
see \cite{Britton:1992pg,Britton:1992xv}. In the smaller mass region ($4\MeV
\lesssim M_\m \lesssim 60\MeV$) Refs.~\cite{Britton:1992pg,Britton:1992xv}
provided the bound based on the change of the number of events in the primary
positron peak located at energies $M_\pi/2$ . Similar bounds were obtained
for the same mixing angle in studies of \textit{kaon} decays
\cite{Yamazaki:1984sj} and for the $\vartheta_\mu^2$ in the decays of both
pions \cite{Bryman:1996xd,Abela:1981nf,Daum:1987bg} and kaons
\cite{Yamazaki:1984sj,Hayano:1982wu}.

\begin{figure}[t]
  \centering
  \includegraphics[width=.5\textwidth]{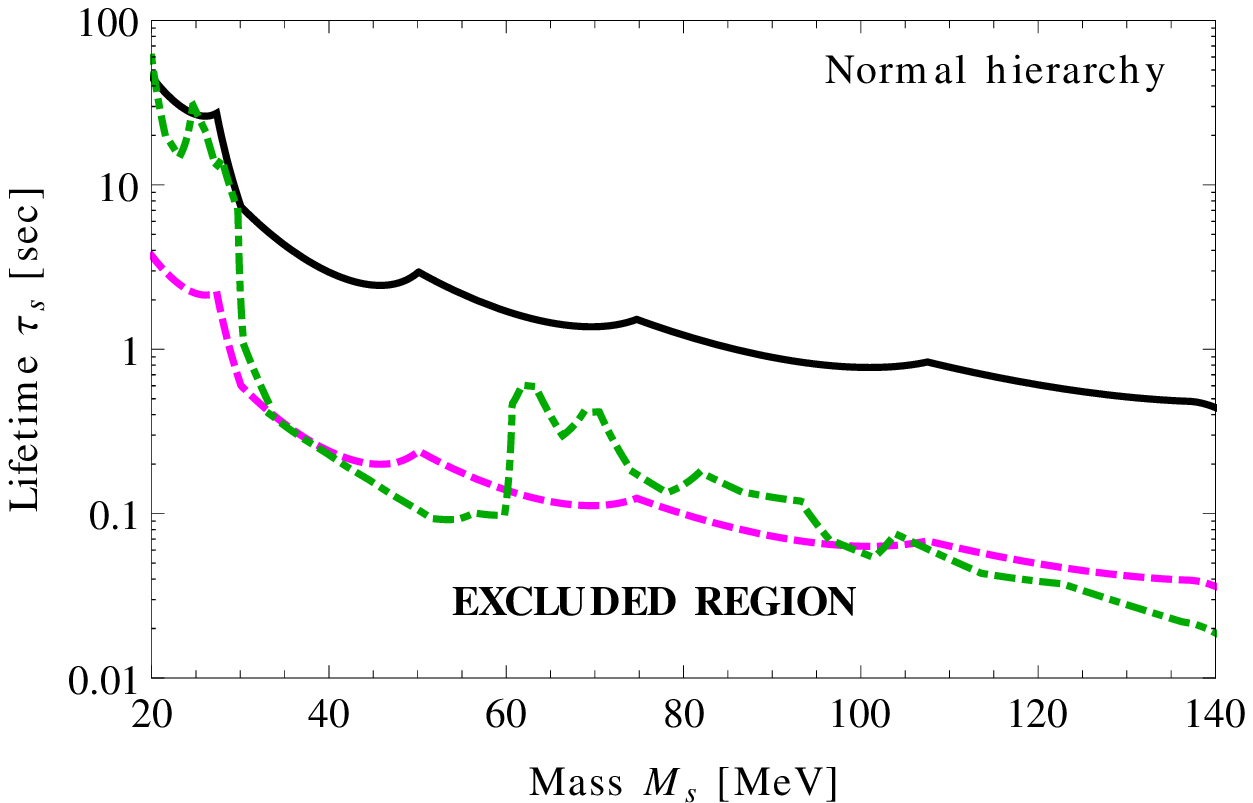}~%
  \includegraphics[width=.5\textwidth]{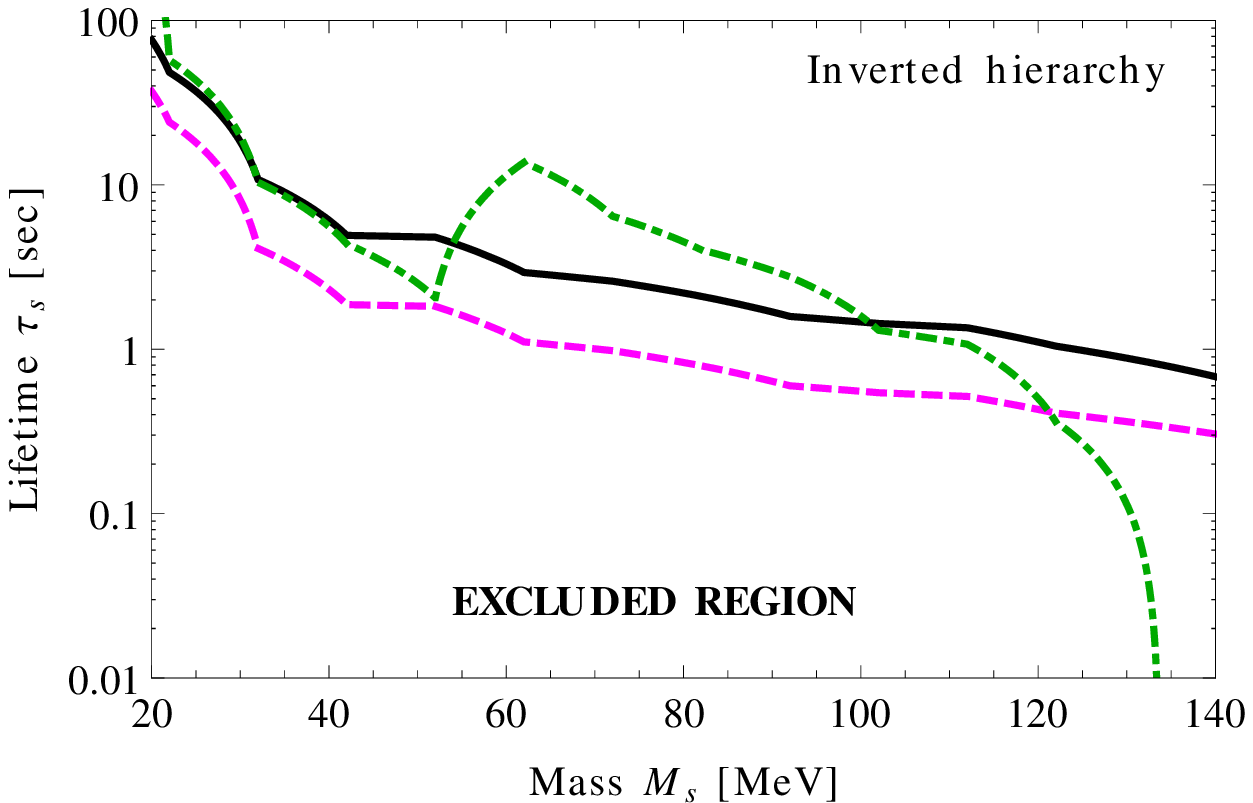}
  \caption{The lower bounds on the lifetime of sterile neutrinos, responsible
    for the mixings between active neutrinos of different flavours in the
    see-saw models~(\protect\ref{eq:nuMSM_Lagrangian}). The bounds are based
    on the combination of negative results of direct experimental searches~\protect\cite{Bernardi:1985ny,Bernardi:1987ek,Britton:1992xv,Aoki:2011vma,Yamazaki:1984sj,Hayano:1982wu,Bryman:1996xd,Abela:1981nf,Daum:1987bg}
    with the neutrino oscillation data~\protect\cite{Schwetz:2011zk}. The neutrino
    oscillation parameters are allowed to vary within their $3\sigma$
    confidence intervals to minimize the lifetime.  The solid black curve is
    based on our reinterpretation of PS191 data \emph{only}, that takes into
    account charged and neutral current contributions (see
    Sec.~\protect\ref{sec:ps191}). The interpretation of the PS191 experiment,
    taking into account only CC interactions (used e.g. in the previous
    works~\protect\cite{Gorbunov:07a,Asaka:2011pb}) is shown in magenta dashed
    line. The bound from peak searches experiments \emph{only}
    \protect\cite{Britton:1992xv,Aoki:2011vma,Yamazaki:1984sj,Hayano:1982wu,Bryman:1996xd,Abela:1981nf,Daum:1987bg}
    is plotted in green dot-dashed line. }
  \label{fig:lifetime-interpretations-peaks}
\end{figure}

The \emph{lower} bound on the sterile neutrino lifetime $\tau_s$ in the
model~(\ref{eq:nuMSM_Lagrangian}), based on the peak search data and neutrino
oscillations is shown in Fig.~\ref{fig:lifetime-interpretations-peaks} by
dot-dashed green lines. The parameters of neutrino mixing matrix are allowed
to vary within their $3\sigma$ limits (to minimize $\tau_s$, while still
keeping the values of all mixing angles compatible with the bounds from direct
experimental searches).

\begin{figure}[ht]
  \centering
  \includegraphics[width=.5\textwidth]{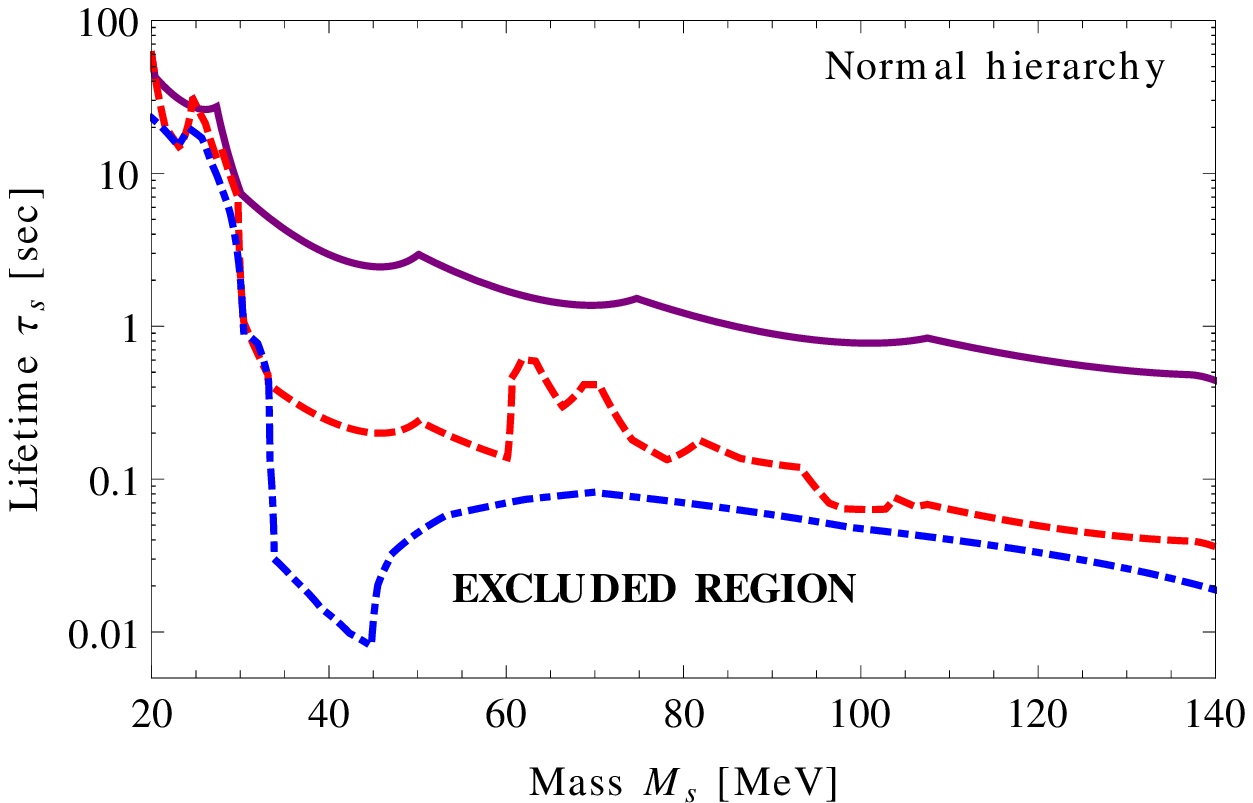}~%
  \includegraphics[width=.5\textwidth]{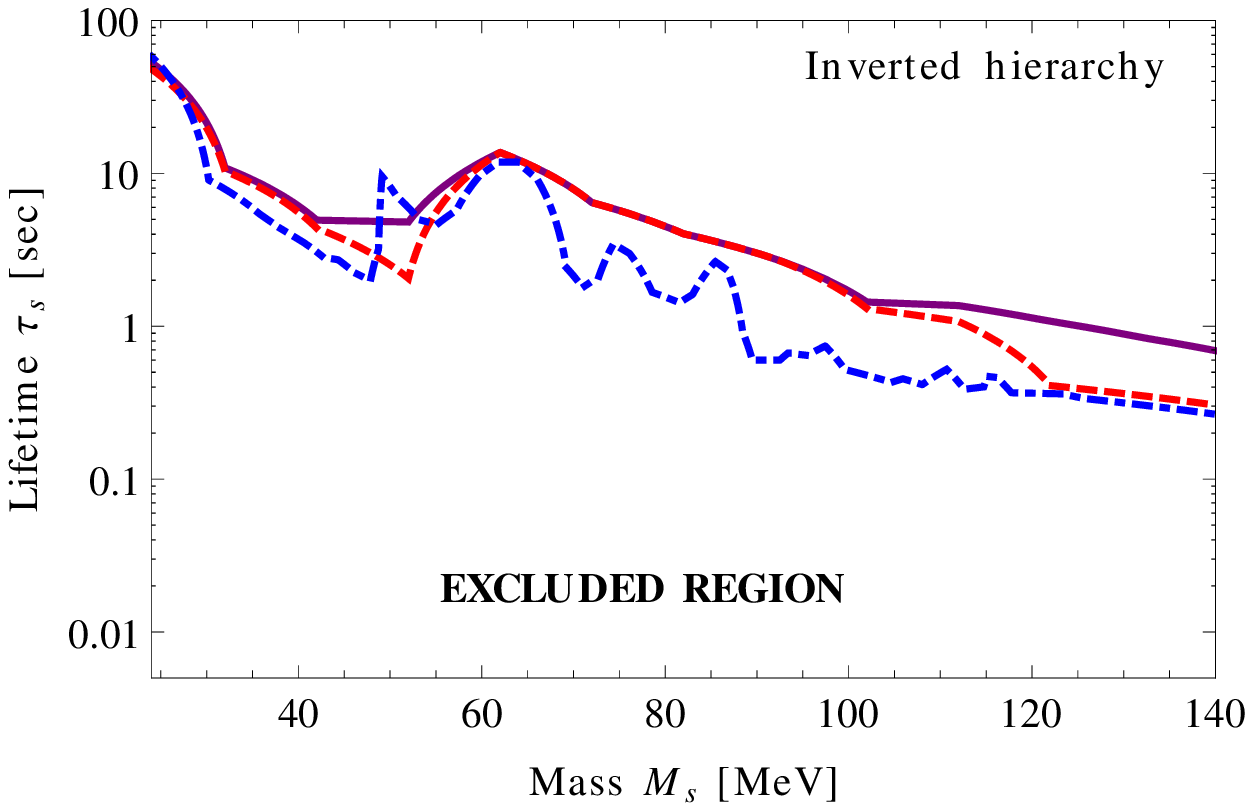}
  \caption{Comparison with the previous bounds on sterile neutrino lifetime in
    the \numsm~\protect\cite{Asaka:2011pb}. The solid purple curves represent
    the results of the present work, obtained by the combination of peak
    searches
    experiments~\protect\cite{Britton:1992xv,Aoki:2011vma,Yamazaki:1984sj,Hayano:1982wu,Bryman:1996xd,Abela:1981nf,Daum:1987bg}
    together with the reanalysis of PS191, that takes into account neutral
    currents (a union of black and green bounds from
    Fig.~\protect\ref{fig:lifetime-interpretations-peaks}).  The red dashed
    curve is based on the combination of the same peak searches with the
    \textit{original} interpretation of PS191 (i.e., with charged current
    interactions only). The blue dot-dashed line is taken
    from~\protect\cite{Asaka:2011pb}. \emph{Notice}, that the results
    of~\protect\cite{Asaka:2011pb} were multiplied by a factor $2$ to account
    for the Majorana nature of the particles (see discussion in
    Sec.~\protect\ref{sec:majorana-vs-dirac}), that was missing therein. The
    difference between the red and blue lines in the case of normal hierarchy
    is explained by wider $3\sigma$ intervals for neutrino oscillation data,
    used in~\protect\cite{Asaka:2011pb}, compared to our work.}
  \label{fig:lifetime-CC+CCNC-prev-studies}
\end{figure}

\begin{figure}[h]
  \centering
  \includegraphics[width=.5\textwidth]{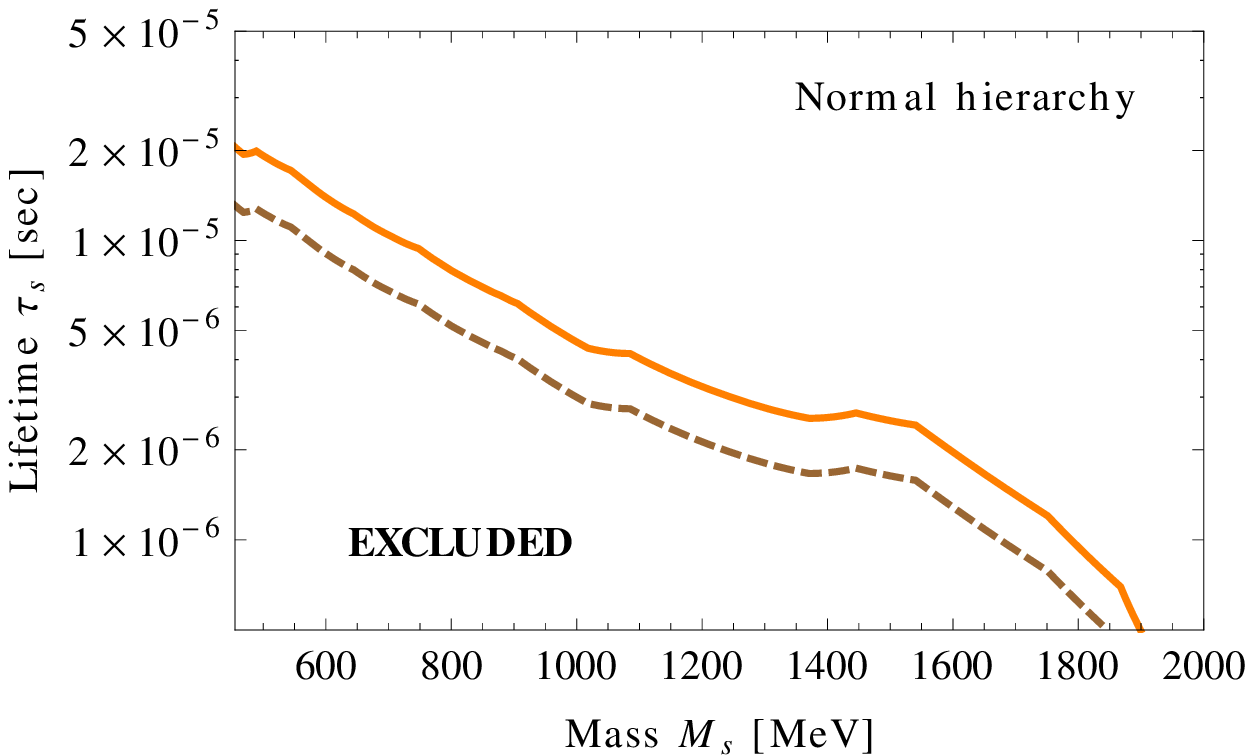}~%
  \includegraphics[width=.5\textwidth]{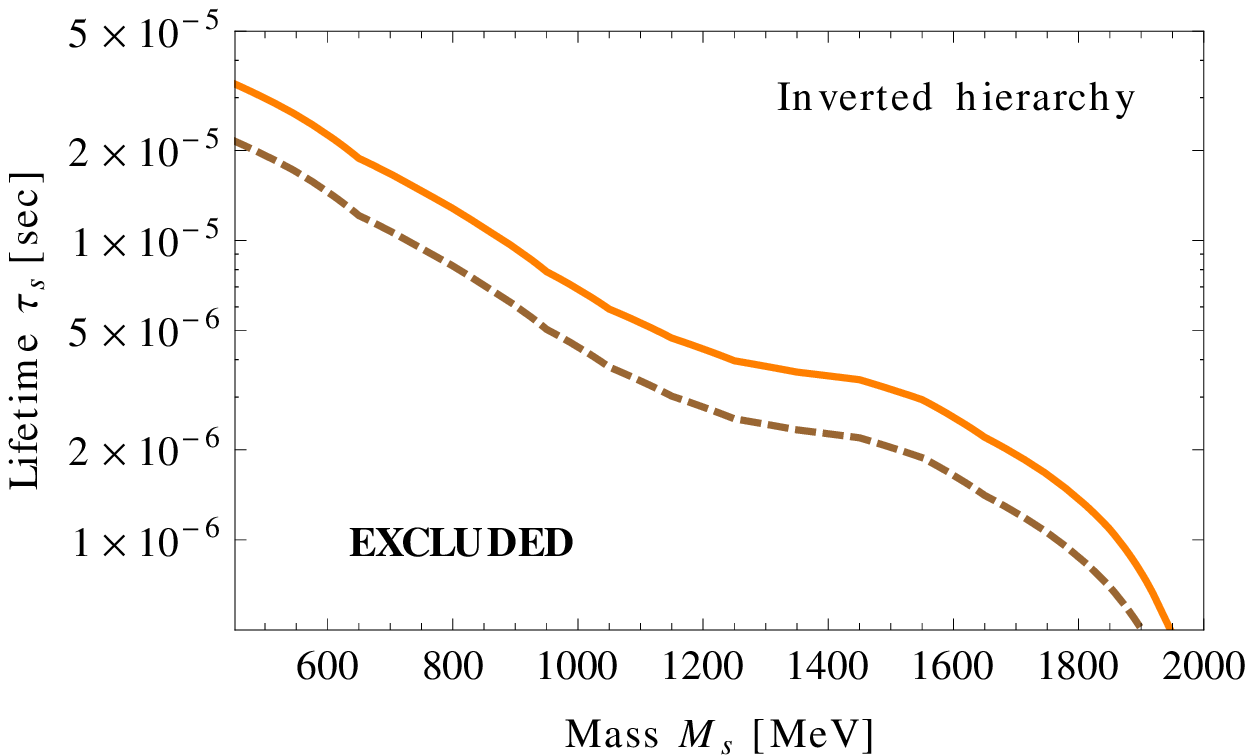}
  \caption{Comparison of the bounds on sterile neutrino lifetime (in the model
    (\protect\ref{eq:nuMSM_Lagrangian})) based on the results of the CHARM
    experiments~\protect\cite{CHARM:1985} \emph{solely} (combined with the neutrino
    oscillation data).  The orange (upper) curves correspond to the model with
    charged and neutral current interactions of sterile neutrinos, the brown
    (lower) -- to the model with charged current interactions only. For
    details, see Sec.~\protect\ref{sec:ps191}. }
  \label{fig:CHARM-only-CCvsCCNC}
\end{figure}

\subsection{Fixed target experiments and neutral currents contribution}
\label{sec:neutr-curr-contr}

The second kind of experiments (``\emph{fixed target experiments}'')
\cite{Bernardi:1987ek,Vaitaitis:1999wq,CHARM:1985} aims to create sterile
neutrinos in decays of mesons and then searches for their decays into pairs of
charged particles. Notice, that the expected signal in this second case is
proportional to $\vartheta_\alpha^4$ or $\vartheta_\alpha^2\vartheta_\beta^2$
(and not to $\vartheta_\alpha^2$ as in the case of peak searches, discussed in
the Section~\ref{sec:peak-searches}).  We will demonstrate below that
 in
the models like~(\ref{eq:nuMSM_Lagrangian}) (and in particular in the \numsm)
the results of some fixed target experiments should be reinterpreted and will
provide stronger bounds than discussed in previous
works~\cite{Gorbunov:07a,Atre:09,Asaka:2011pb} (see
also~\cite{Kusenko:2004qc}).

\subsection{Reinterpretation of the PS191 and CHARM experiments}
\label{sec:ps191}

The experiment \textbf{PS191} at CERN was a ``fixed target'' type of experiment
described above~\cite{Bernardi:1985ny,Bernardi:1987ek}. In searches for
sterile neutrinos lighter than the pion $M_\m < M_\pi$, the pair of charged
particles that were searched for in the neutrino decay comprised mostly of
electron and positron:
\begin{equation}
  \label{eq:meson-decay-chain-electron}
  \begin{array}{rl}
    \pi^+/K^+ \to e^+\; + &N \\
    &\hookrightarrow e^+ \,e^- \nu_\alpha\;,
  \end{array}
\end{equation}
where $N$ is a sterile neutrino with the mass $M_\m$. The first reaction in
the chain is solely due to the \emph{charged-current} (CC) interaction, and
its rate is proportional to the
$\vartheta_e^2$.

If sterile neutrinos interact through both \emph{charged and neutral currents}
(CC+NC) as it is the case in the models with the see-saw
Lagrangian~(\ref{eq:nuMSM_Lagrangian}), any of three active-neutrino flavours
may appear in the decay of $N$ in (\ref{eq:meson-decay-chain-electron}). The
decay widths are \cite{Shrock:80}:
\begin{equation}
\label{eq:9}
  \Gamma(N\rightarrow e^+ e^- \nu_\alpha) =  c_\alpha \vartheta_\alpha^2 \frac{G_F^2 M_\m^5}{96\pi^3}, 
\end{equation}
with the following definition\footnote{Note that in the Ref. \cite{Astier:2001ck} there is a typo in the expression for $c_\tau$ (Eq. (2)).}
\begin{equation}
\label{eq:c_alpha-definition}
c_e=\frac{1+4\sin^2\theta_W + 8 \sin^4\theta_W}{4},~~ c_\mu=c_\tau=\frac{1-4\sin^2\theta_W + 8
  \sin^4 \theta_W}{4}, 
\end{equation}  
and $\theta_W$ is the Weinberg's angle so that $\sin^2\theta_W \approx 0.231$
and $c_e\approx 0.59,~c_{\mu(\tau)}\approx 0.13$.  Therefore, the total number
of events inside the detector that registers electron-positron pairs would be
proportional to the combination of mixing angles $\vartheta_e^2 \times (\sum
c_\alpha \vartheta_\alpha^2)$.

However, the model employed in the interpretation of the PS191
experiment~\cite{Bernardi:1985ny,Bernardi:1987ek} was different, as has
already been pointed in~\cite{Kusenko:2004qc}.  In the original analysis
it was assumed that sterile neutrino interacts \textit{only via charged
  currents}, but not through neutral currents. In our language it means that
$c_e=1,c_{\mu(\tau)}=0$ was used instead of the values
(\ref{eq:c_alpha-definition})\footnote{Model described in
  \cite{Bernardi:1985ny,Bernardi:1987ek} contains only one Dirac neutrino,
  while in the $\nu$MSM we have two Majorana fermions. Therefore actually
  $c_e=1/2$ in the original model. For details see
  Sec.~\ref{sec:majorana-vs-dirac}}. As was noticed above, the probability of
meson decay into sterile neutrino does not alter if we exclude the
neutral-current interaction, and therefore the total number of events with the
electron-positron pair would be proportional to $\vartheta_e^2 \times
\vartheta_e^2$.

Therefore if we denote the bounds listed in
\cite{Bernardi:1985ny,Bernardi:1987ek} as $\vartheta_e^4 \leq
\vartheta_{e(exp)}^4$, then the bound for the \numsm takes form
\begin{equation}
\label{eq:PS191-reanalysis-bound}
\vartheta_e^2 ~\left(\sum_{\alpha = \{e,\mu,\tau\}} c_\alpha \,\vartheta_\alpha^2\right) \leq \vartheta_{e(exp)}^4\;.
\end{equation} 
Similar bounds can be extracted from the reanalysis of meson decays into
\textit{muon} and sterile neutrino, that leads to replacement $e\to\mu$ in
(\ref{eq:PS191-reanalysis-bound}).  As a result, the reinterpretation of the
results of the PS191 experiment in combination with neutrino oscillation data
produces up to an order of magnitude \emph{stronger} bounds on lifetime than in the
previous works (see Figs.~\ref{fig:lifetime-interpretations-peaks}
and~\ref{fig:lifetime-CC+CCNC-prev-studies}).

Similarly, the CHARM experiment \cite{CHARM:1985} provided bounds on the
mixing angles of sterile neutrinos in the mass range $0.5\gev \lesssim M_\m
\lesssim 2\gev$. In the original analysis NC contributions \emph{were
  neglected}. Therefore, to apply the results of this experiment to the case
of the \numsm, we reanalyzed the data as described above. In
Fig.~\ref{fig:CHARM-only-CCvsCCNC} we compare lifetime bounds coming from the
CHARM experiment solely for CC and CC+NC interactions of sterile
neutrinos. The difference in this case is about a factor of 2.\footnote{In the
  case of the PS191 experiment, when using CC only for masses below the mass
  of pion suppression of the $\vartheta_e^2$ mixing angle due to neutrino
  oscillations meant that instead of $\vartheta_e^2$ bounds the lifetime is
  defined by the (much weaker) $\vartheta_\mu^2$ bounds. That led to the
  significant relaxation of the lower bound on the lifetime. If NC were taken
  into account, this was not possible anymore and therefore the lower bound on
  sterile neutrino lifetime became stronger by as much as the order on
  magnitude (black vs. magenta curve on the left panel in
  Fig.~\ref{fig:lifetime-interpretations-peaks}.  In case of the CHARM
  experiment, both $\vartheta_e^2$ and $\vartheta_\mu^2$ are strongly
  constrained and switching from one constraint to another makes (numerically)
  much smaller difference.}

\subsection{A note on Majorana vs Dirac neutrinos}
\label{sec:majorana-vs-dirac}

For completeness we briefly discuss the difference in interpreting
experimental results for \emph{Majorana vs. Dirac sterile neutrinos}. Similar
discussion can be found e.g.\ in~\cite{Gorbunov:07a}.  When interpreting the
experimental results one should take into account that in present work we
consider \emph{two Majorana sterile neutrinos}, while the experimental papers
often phrase their bounds in terms of the mixing with a single \emph{Dirac}
neutrino, that we will denote $U_{\alpha}^2$.  In the \numsm twice more
sterile neutrinos are produced per single reaction (because there are two
sterile species -- $N_2$ and $N_3$), and, owing to their Majorana nature, each
sterile neutrino decays twice faster (additional charge-conjugated decay modes
are present).  Notice, that the mass splitting between between two sterile
states $N_2,N_3$ is small $|M_2 - M_3|\ll \frac12 (M_2+M_3) = M_s$ and once
born, the states oscillate fast into each other. Averaging over many
oscillations can be accounted for by an extra factor $\frac12$ in the number
of $N_2$ and $N_3$ species.  Therefore, for fixed target experiments one gets the
same number of the detector events involving one Dirac sterile neutrino as one
gets in the \numsm if $(\vartheta_{\alpha2}^2+\vartheta_{\alpha3}^2)^2 =
U_\alpha^4$. That is, one should identify $2\vartheta_\alpha^2$ with the
measured $U_\alpha^2$ (recall~(\ref{eq:19}) that $\vartheta^2_\alpha =
\frac12(\vartheta^2_{\alpha2}+\vartheta^2_{\alpha3})$).  In the case of peak
searches, the bound $U_\alpha^2$ should be interpreted in the \numsm as
$\vartheta_{\alpha,2}^2 + \vartheta_{\alpha,3}^2 \le U_\alpha^2$, as
production of \emph{any} state $N_2$ or $N_3$ contributes to the number of
events in the secondary peak, i.e. again $2\vartheta_\alpha^2$ should be
identified with $U_\alpha^2$.  Notice, that this factor 2 is missing
in~\cite{Asaka:2011pb}.

\begin{figure}[t]
  \centering \subfloat[Bound on the combination $\vartheta_e\sqrt{\strut
    \sum\limits_\alpha c_\alpha \vartheta_\alpha^2}$]%
  [Bound on the combination $\vartheta_e\sqrt{\strut \sum c_\alpha
    \vartheta_\alpha^2}$]%
  { \includegraphics[width=.5\textwidth]{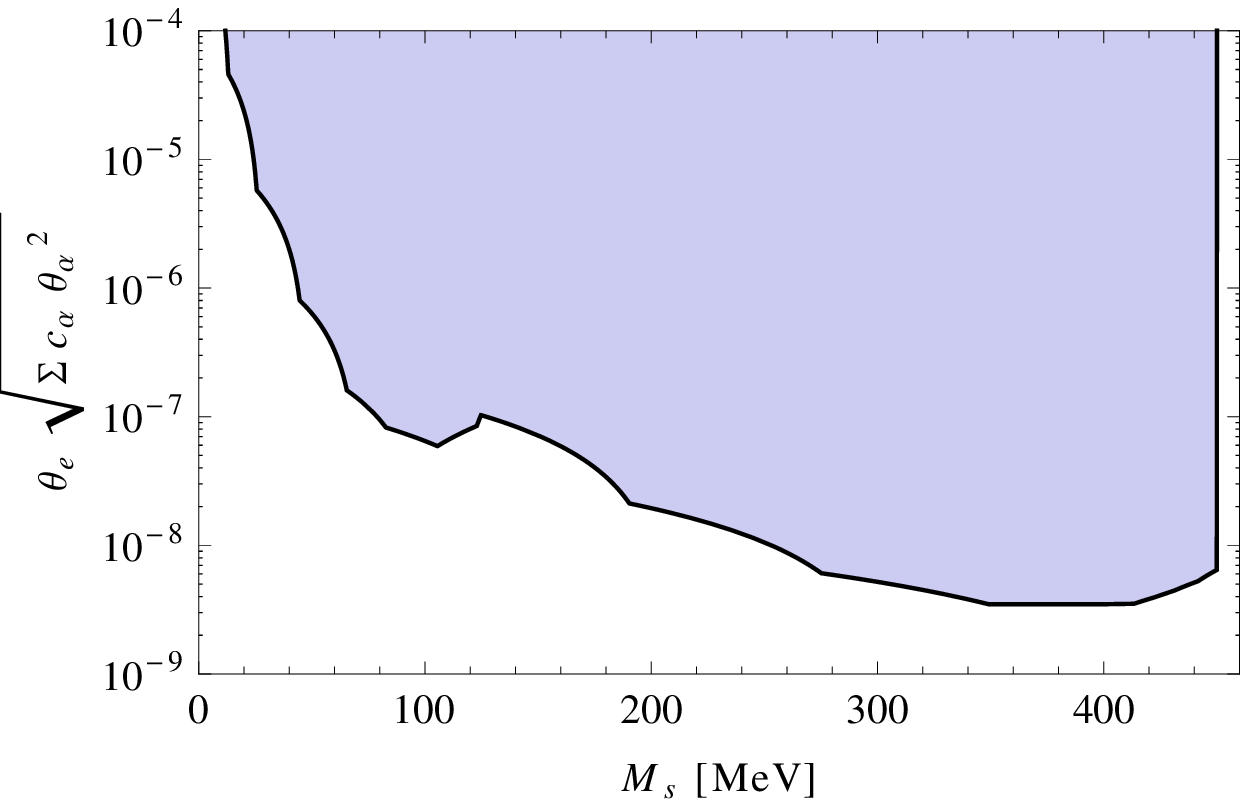}}~%
  \subfloat[Bound on the combination $\vartheta_e\sqrt{\strut
    \sum\limits_\alpha c_\alpha \vartheta_\alpha^2}$]%
  [Bound on the combination $\vartheta_\mu\sqrt{\strut \sum c_\alpha
    \vartheta_\alpha^2}$]%
  { \includegraphics[width=.5\textwidth]{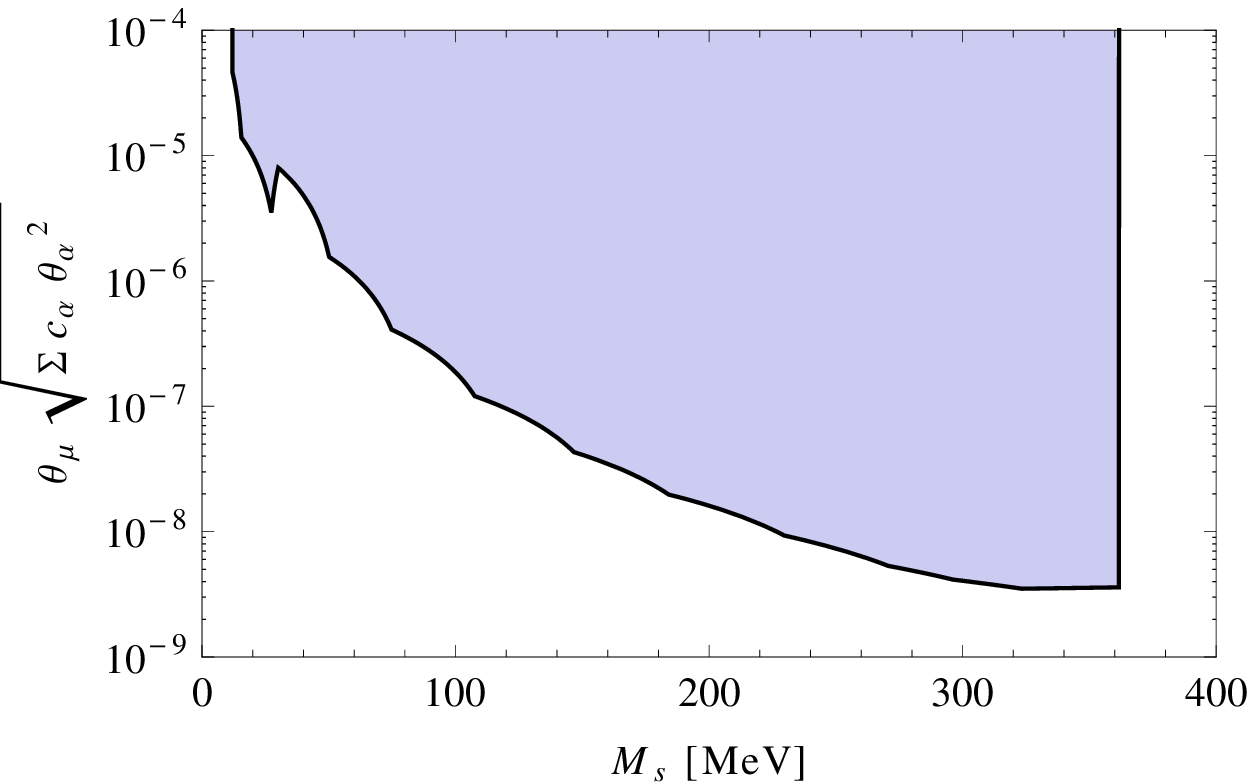}}
  \caption{Direct accelerator bounds on the combination of active-sterile
    neutrino mixing angles, resulting from the reanalysis of the PS191
    experiment~\protect\cite{Bernardi:1985ny,Bernardi:1987ek}, taking into account decays of sterile neutrino through both
    charged and neutral currents and their Majorana nature. The shaded region
    is excluded.  The case, analyzed in the original
    works~\protect\cite{Bernardi:1985ny,Bernardi:1987ek} (decay of sterile neutrino
    through the charged current only) corresponds to the choice $c_e = 1$,
    $c_\mu=c_\tau = 0$, for details, see Sec.~\protect\ref{sec:ps191}.  We plot the
    bounds for \emph{two Majorana} neutrinos (as in
    Fig.~\protect\ref{fig:mixing_constraints}) while in the original
    works~\protect\cite{Bernardi:1985ny,Bernardi:1987ek} a single Dirac neutrino was
    analyzed.}
  \label{fig:PS191-bounds}
\end{figure}
\begin{figure}[h]
  \centering
  \includegraphics[width=.5\textwidth]{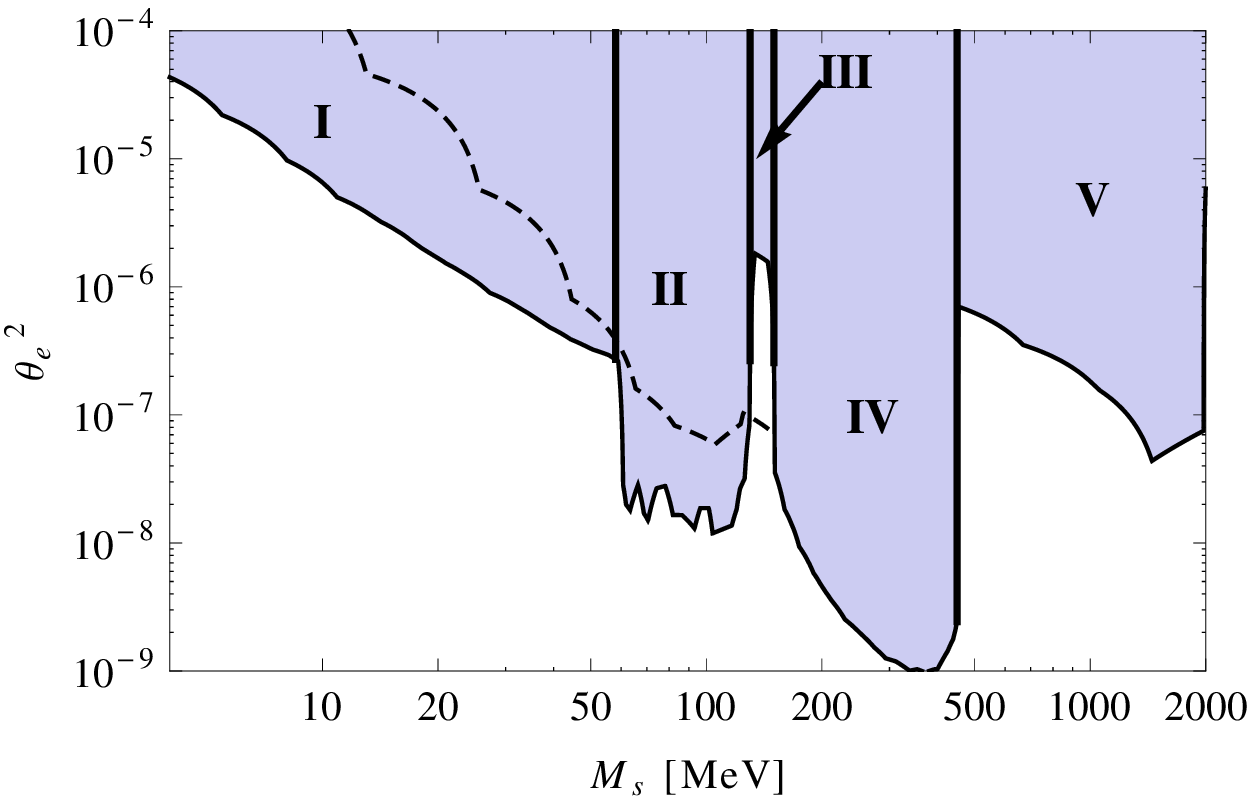}%
  \includegraphics[width=.5\textwidth]{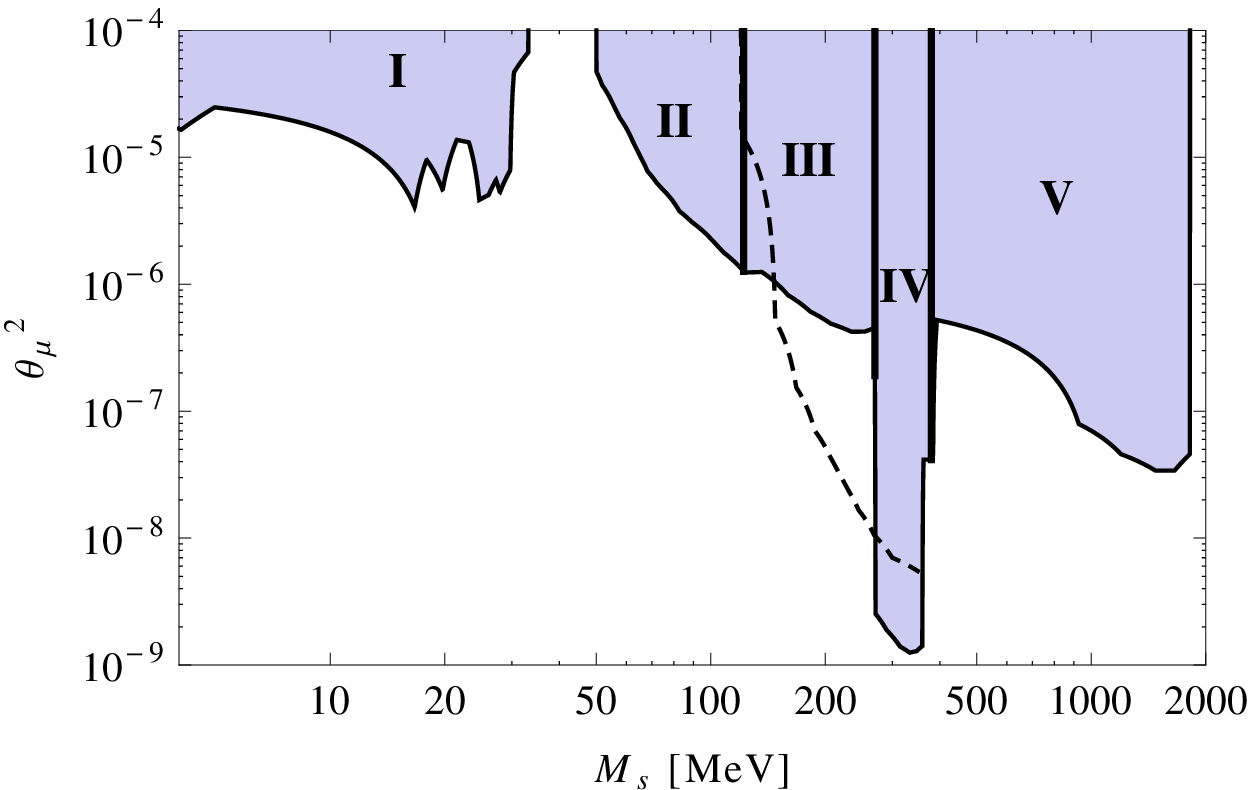}%
  \caption{Direct accelerator bounds on the mixing angles. \textbf{Left
      panel:} $\vartheta_e^2$ bounds, taken from~\protect\cite{Britton:1992xv} (region
    \Rmnum{1}),~\protect\cite{Aoki:2011vma} (region \Rmnum{2}),~\protect\cite{Yamazaki:1984sj}
    (region \Rmnum{3}), \protect\cite{Bernardi:1987ek,Levy} (region \Rmnum{4}) and
    \protect\cite{CHARM:1985} (region \Rmnum{5}). \textbf{Right panel:}
    $\vartheta_\mu^2$ bounds, taken from
    \protect\cite{Bryman:1996xd,Abela:1981nf,Daum:1987bg} (region \Rmnum{1}),
    \protect\cite{Yamazaki:1984sj}(region \Rmnum{2}), \protect\cite{Hayano:1982wu} (region
    \Rmnum{3}), \protect\cite{Bernardi:1987ek}(region \Rmnum{4}) and
    \protect\cite{Vaitaitis:1999wq} (region \Rmnum{5}). The shaded regions are ruled
    out by the experimental findings.
 Dashed curves indicate mixing angle
    bounds given by original interpretation of PS191 experiment, but we
    \textit{do not} use them to derive our final results, as explained in
    Sec.~\protect\ref{sec:ps191}. The bounds are shown for the Majorana neutrino and
    are therefore two times \textit{stronger} (see
    Section~\protect\ref{sec:majorana-vs-dirac}), while in the original
    works~\protect\cite{Bernardi:1985ny,Bernardi:1987ek} a single Dirac neutrino has
    been considered.}
  \label{fig:mixing_constraints}
\end{figure}

\section{Results}
\label{sec:discussion}

In this Section we summarize our results: the upper bound on the (combination
of) mixing angles of sterile and active neutrinos in the see-saw
models~(\ref{eq:lifetime_expression}) in the range 10~MeV -- 2~GeV and the
lower bound on sterile neutrino lifetime, obtained in combination of these
bounds with constraints, coming from neutrino oscillation data.

\begin{figure}[htp]
  \centering %
  \subfloat
  [Normal hierarchy, mass range $10\mev - 2\gev$]
  {\includegraphics[width=.5\textwidth]{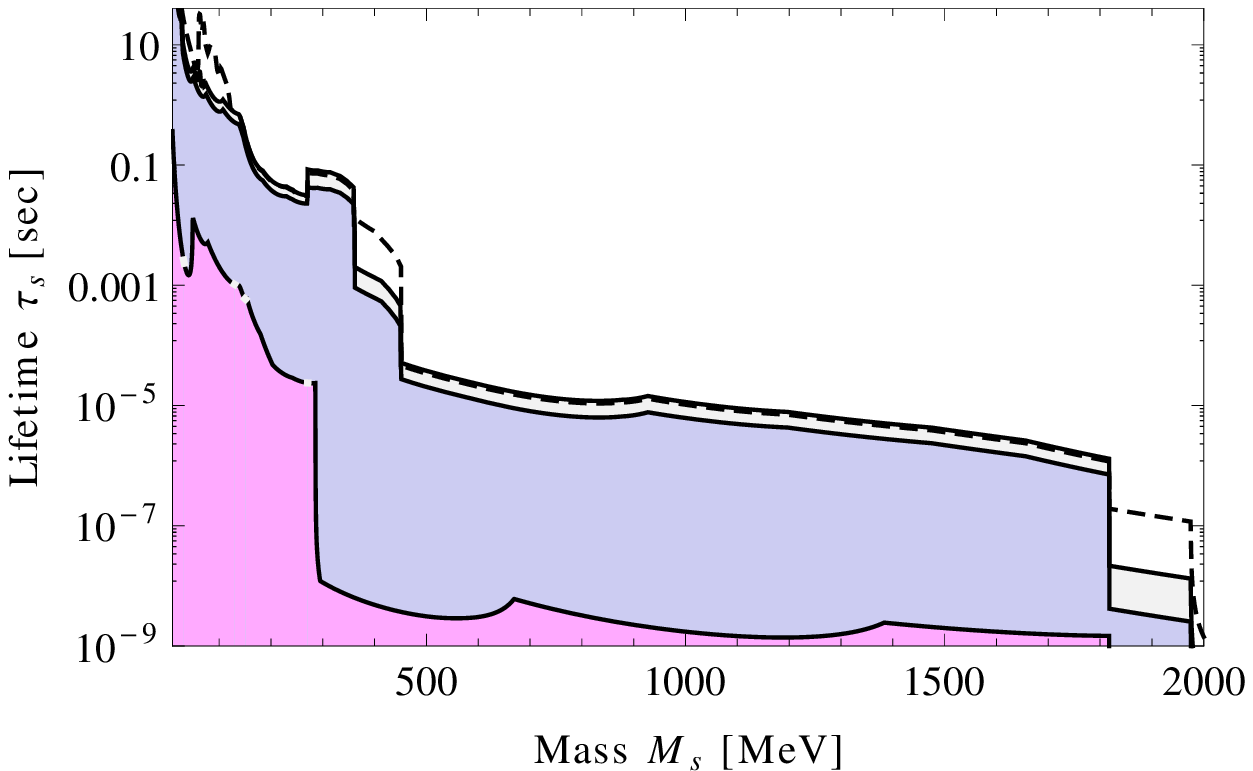}}%
  ~\subfloat
  [Normal hierarchy, zoom at the mass range $10\mev - 140\mev$] %
  {\includegraphics[width=.5\textwidth]{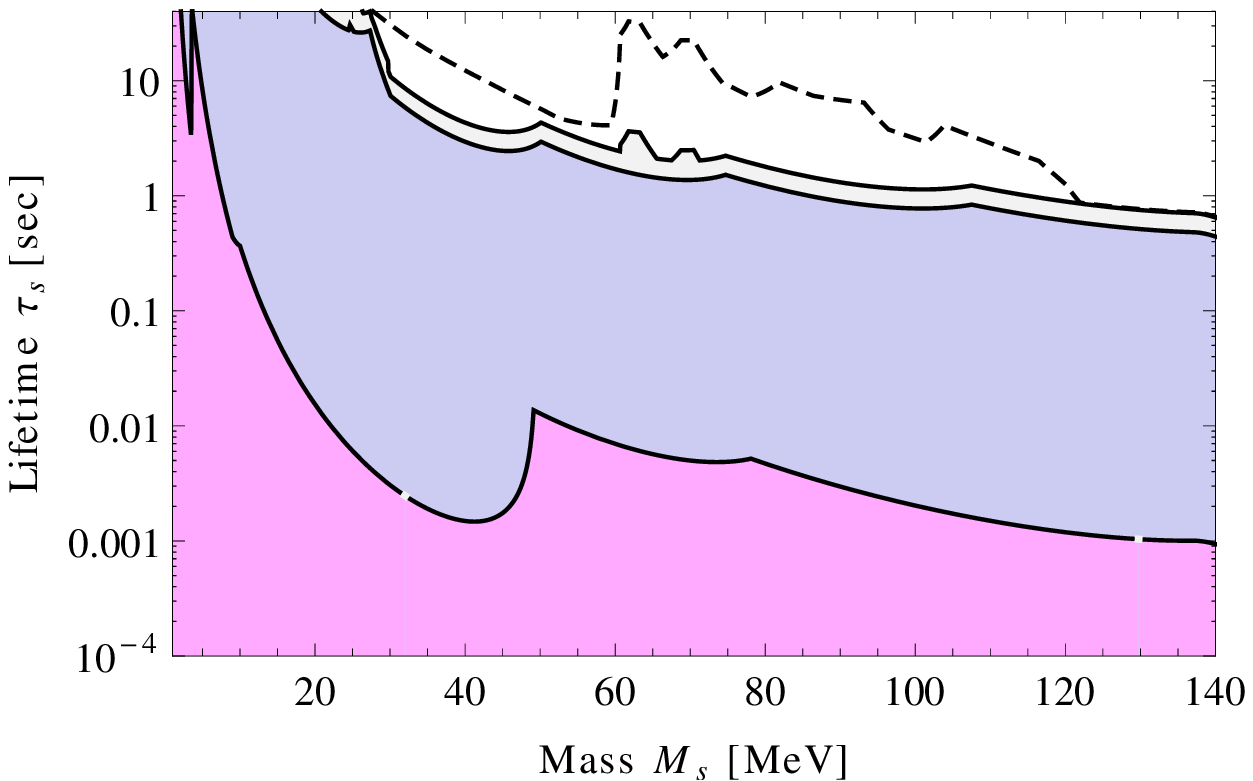}}\\
  \subfloat%
  [Inverted hierarchy, mass range $10\mev- 2\gev$] {
    \includegraphics[width=.5\textwidth]{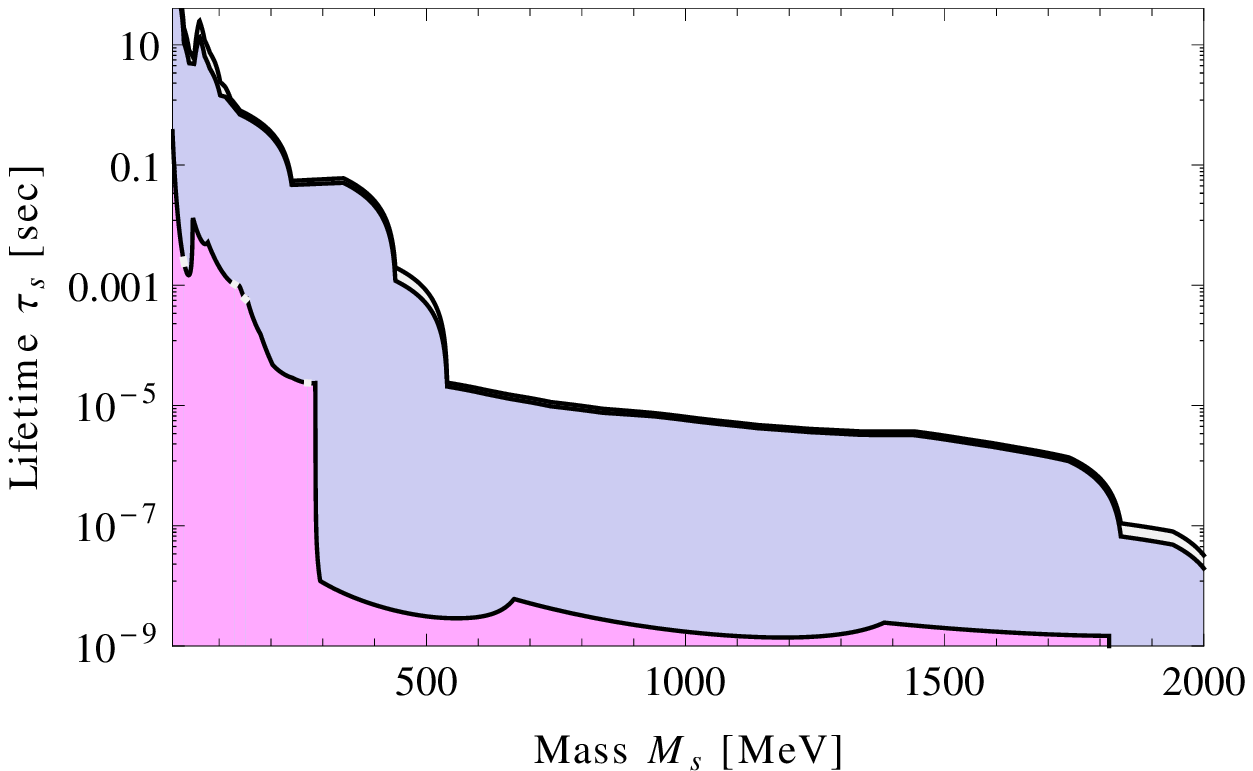}
  }%
  \subfloat
  [Inverted hierarchy, zoom at the mass range $10\mev - 140\mev$] {
    \includegraphics[width=.5\textwidth]{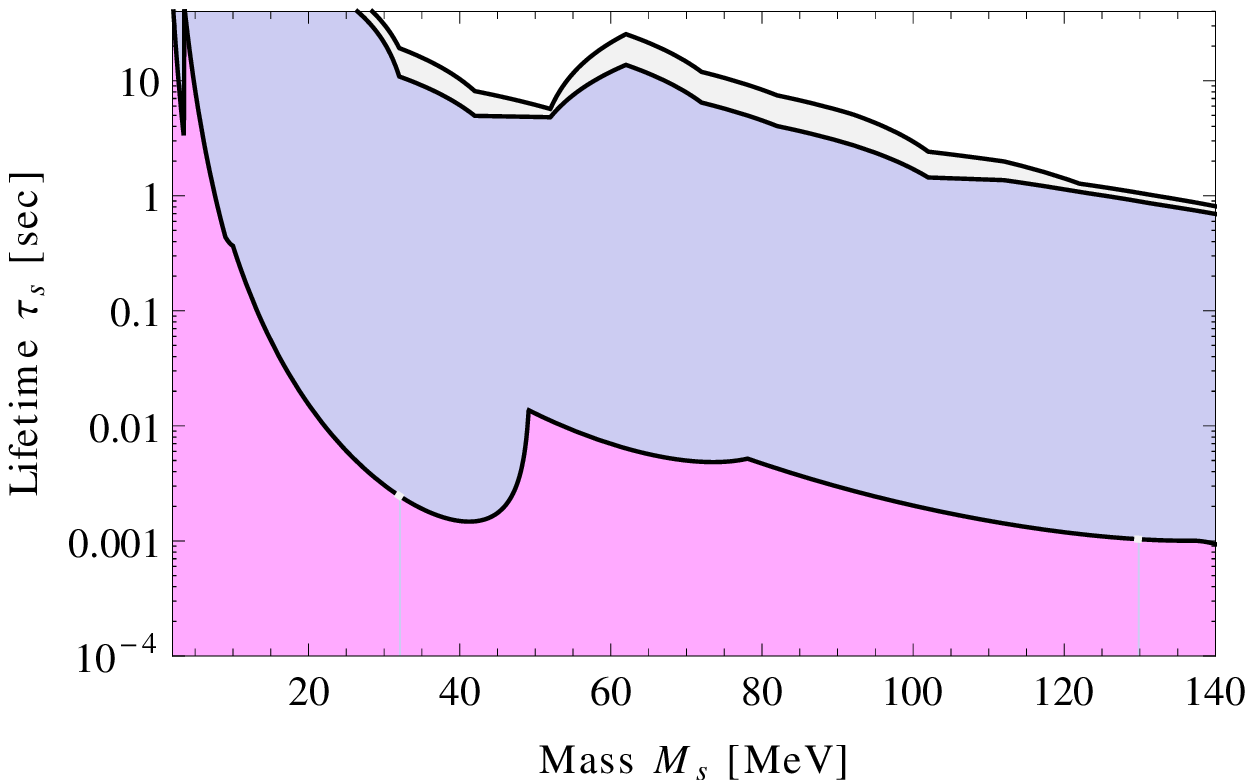}%
    }
    \caption{The resulting lower bounds on sterile neutrino lifetime $\tau_s$
      as a function of their mass, obtained by requiring that two Majorana
      sterile neutrinos are responsible for neutrino oscillations and their
      parameters do not contradict the negative results of direct experimental
      searches. In all figures the upper curve comes from using of the best
      fit neutrino oscillation parameters, the middle one -- from their
      variation within the $3\sigma$ limits, and the lower one does not take
      into account neutrino oscillation data and puts all three mixing angles
      equal to their direct experimental bounds. The dashed line for NH
        corresponds to the best-fit values of PMNS parameters with
        $\theta_{13} = 0$ and shows how much the bounds on the lifetime relax
        for non-zero value of $\theta_{13}$ (see text,
        Section~\protect\ref{sec:discussion} for discussion). }
\label{fig:lifetime_result}
\end{figure}

\subsection{Bounds on the mixing angles of sterile neutrinos}
\label{sec:bounds-mixing}

For the models~(\ref{eq:nuMSM_Lagrangian}) (two Majorana sterile neutrinos,
interacting through both charged and neutral interactions), the compilation of
constraints on various combinations of active-sterile mixing angles
($\vartheta_e^2$, $\vartheta_\mu^2$, $\vartheta_e \sqrt{\sum c_\alpha
  \vartheta_\alpha^2}$, $\vartheta_\mu \sqrt{\sum c_\alpha
  \vartheta_\alpha^2}$) that we used in this work are plotted in
Figs.~\ref{fig:PS191-bounds} and~\ref{fig:mixing_constraints}.\footnote{Notice
  that in the published results of the PS191 experiment~\cite{Bernardi:1987ek}
  bounds are given up to $M_\m=400\MeV$. We extend these bounds up to
  $450\MeV$, using the PhD Thesis of J.-M. Levy \cite{Levy}.}

\subsection{The lower bound on the lifetime of sterile neutrinos}
\label{sec:bounds-lifetime}

The result of the
Sections~\ref{sec:normal-hierarchy}--\ref{sec:inverted-hierarchy}, combined
with these experimental bounds can be translated into the \emph{lower} limits
on the lifetime of sterile neutrinos. These results are presented in
Figs.~\vref{fig:lifetime_result}.  Additionally, we plot the lifetime bounds
for the best-fit values of the PMNS parameters yet with $\theta_{13}=0$ (as
used e.g. in~\cite{Gorbunov:07a,Canetti:10a}). For normal hierarchy we see
that our bounds with $\theta_{13} \neq 0$ are relaxed by as much as the order
of magnitude at some masses, compared to $\theta_{13}=0$ case. The difference
for IH is not so pronounced. \emph{Notice}, that the bounds
of~\cite{Gorbunov:07a,Canetti:10a,Asaka:2011pb} were different from what we
show as dashed line in Fig.~\ref{fig:lifetime_result} because of ignoring the
neutral current contributions to the results of PS191 experiment (for details
see discussion in Section~\ref{sec:experiments} and
Figs.~\ref{fig:lifetime-interpretations-peaks},~\ref{fig:lifetime-CC+CCNC-prev-studies}).

\section{Discussion}
\label{sec:conclusion}

In this work we have investigated experimental restrictions on the parameters
of the see-saw Lagrangian in the case when two sterile neutrinos with the
masses between $\sim 10$~MeV and $2$~GeV are responsible for neutrino
oscillations. Combined with the results of the direct experimental searches,
the neutrino oscillation data provide stringent lower bounds on their
lifetime, $\tau_s$ and allows to determine both \emph{maximum} and
\emph{minimum} values of the mixing angles~$\vartheta_\alpha^2$.

We have reinterpreted the results of the PS191
experiment~\cite{Bernardi:1985ny,Bernardi:1987ek}, following
\cite{Kusenko:2004qc}, by taking into account not only charged, but also
neutral-current interactions (as both of these are present in the Type~I
see-saw Lagrangian). Our results demonstrated that below the mass of the pion
the fixed target experiments ($\vartheta^4$ experiments) provide \emph{stronger}
restrictions than the peak search experiments ($\vartheta^2$ experiments) in
case of \textit{normal hierarchy}.  In \textit{inverted hierarchy} the
reanalysis of the PS191 experiment turns out to be very important as
well. In the original analysis of the CHARM experiment
\cite{Vaitaitis:1999wq} neutral-current contributions were neglected as
well and we have reinterpreted these results in a similar way to
PS191. The final results are presented in Figs.~\ref{fig:lifetime_result}.

Future experiments (for example, NA62 in
CERN~\footnote{\url{http://na62.web.cern.ch/NA62}}, LBNE experiment in
FNAL\footnote{\url{http://lbne.fnal.gov}} or upgraded LHCb experiment) have a
great potential of discovering light neutral leptons of the \numsm or
significantly improving the bounds on their parameters (see discussion
in~\cite{Akiri:2011dv} and~\cite{Abazajian:2012ys}). Due to the strong
  suppression of the mixing angles $\vartheta_e^2$ in the case of NH and
  $\vartheta_\mu^2$ in the case of IH, the peak searches in
  the kaon decays (such as e.g.~\cite{Shaykhiev:2011zz}) may miss the sterile
  neutrino (cf.~\cite{Asaka:2011pb}).\footnote{ GeV-scale sterile
  neutrinos in the models with extended Higgs sector~\cite{Kusenko:06a} can be
  searched at the LHC~\cite{Shoemaker:2010fg}.}

\begin{figure}[h]
\centering
\includegraphics[width=.49\textwidth]{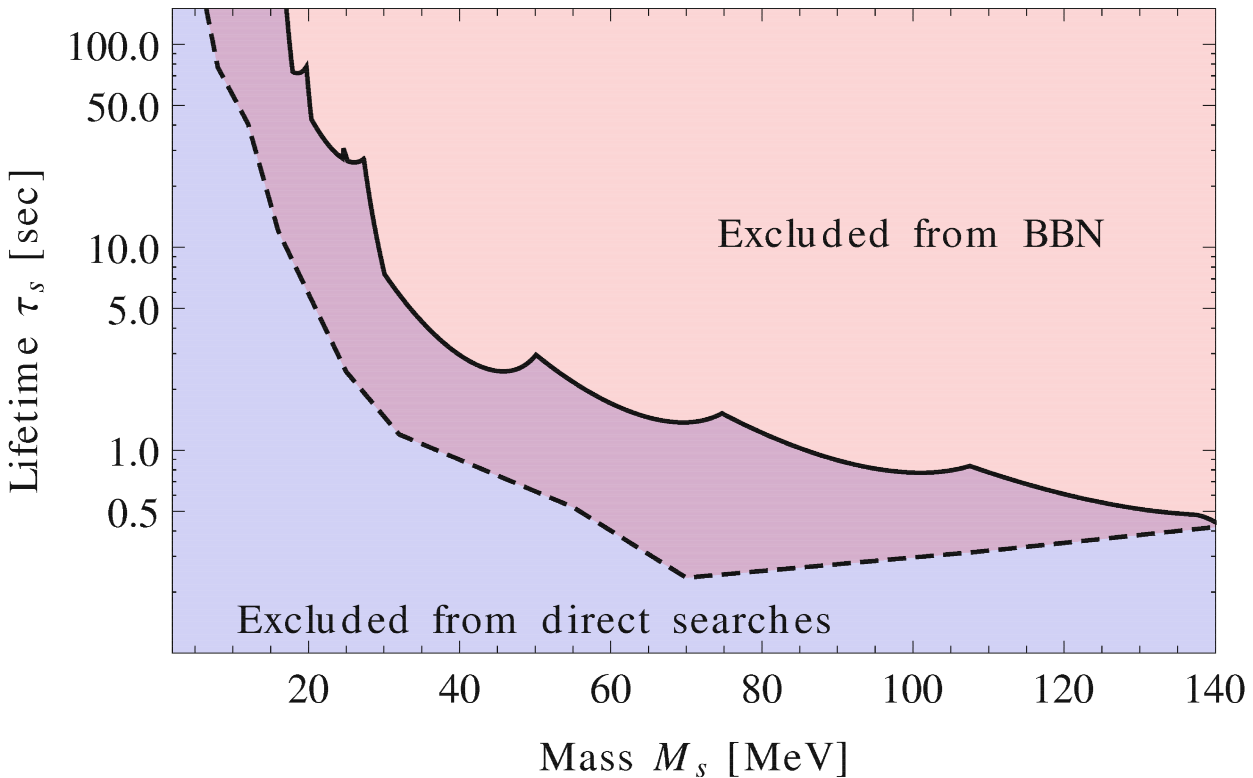}~
\includegraphics[width=.49\textwidth]{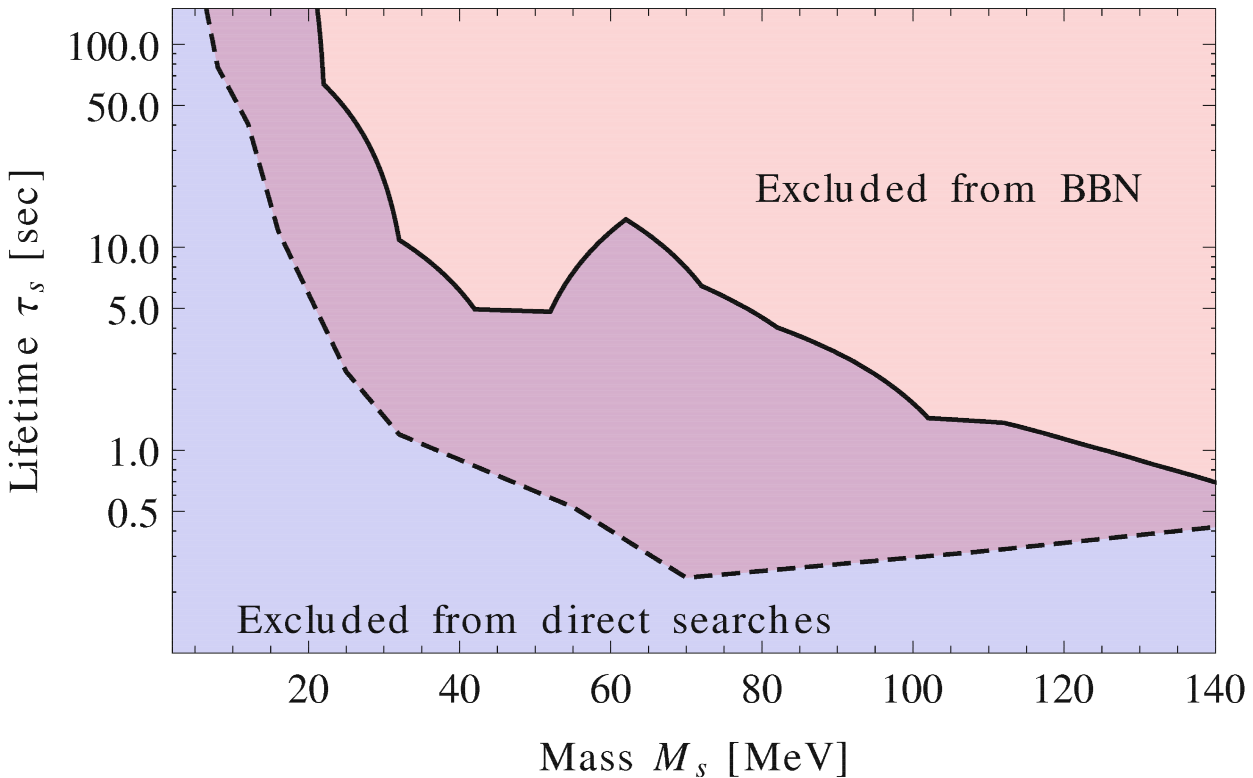}
\caption{Experimental \emph{lower} bounds on the lifetime of sterile
  neutrinos (solid line), combined with the \emph{upper} bounds coming from
  the BBN (dashed line) \protect\cite{Ruchayskiy:2012si}. \textbf{Left panel:}
  normal hierarchy, \textbf{right panel:} inverted hierarchy. Shown are
  $3\sigma$ lower bounds from the Fig.~\protect\ref{fig:lifetime_result} and
  the weakest bound from primordial nucleosynthesis (based on the CMB
  measurements of primordial Helium abundance, yielding $Y_p=0.445$ at the
  $3\sigma$ level~\protect\cite{Dunkley:2010ge,WMAP7}, see text for details).
}
\label{fig:lifetime-BBN}
\end{figure}

Finally, sterile neutrinos with the masses in MeV--GeV region and small mixing
angles can play significant role in the early Universe (for a review
see~\cite{Boyarsky:09a}). Due to their small mixing angles, the lifetime of
sterile neutrinos can be long enough for them to be present in primordial
plasma at $\sim \mev$ temperatures, affecting the Big Bang nucleosynthesis
(BBN) (for more details and references,
see~\cite{Dolgov:00a,Dolgov:00b,Ruchayskiy:2012si}). Comparison of the
$3\sigma$ lower bounds from the direct searches with the $3\sigma$
\emph{upper} bounds from BBN on the lifetime of sterile neutrinos (based
on~\cite{Ruchayskiy:2012si}) is presented in
Fig.~\ref{fig:lifetime-BBN}. There are no allowed values of sterile neutrinos
lifetimes for $10\mev \lesssim M_s < 140\MeV$ for both types of hierarchies
(i.e. the upper bound is \emph{smaller} than the lower bound, see the purple
double-shaded region in Fig.~\ref{fig:lifetime-BBN}). This conclusion is based
on the weakest BBN bound, based on the primordial Helium-4 abundance as
determined from CMB observations~\cite{Dunkley:2010ge,WMAP7}, which presently
has large error bars. The astrophysical measurements
(e.g.~\cite{Aver:2011bw,Izotov:2010ca}) provide a much tighter bounds on the
Helium-4 abundances, further lowering the upper bound from BBN by as much as
the order of magnitude (see~\cite{Ruchayskiy:2012si} for discussion). It
should be stressed that adopting the BBN bounds from the previous
works~\cite{Dolgov:00a,Dolgov:00b} would lead to essentially the same
conclusion.

Our conclusion differs from that of \cite{Asaka:2011pb}, where it was argued
that in the case of normal hierarchy there is an open region of parameter
space compatible with both direct experimental searches and BBN bounds
of~\cite{Dolgov:00b}.  This difference comes mainly from the reanalysis of the
data of PS191 experiment that was carried out in the present paper (as
illustrated in Fig.~\ref{fig:lifetime-CC+CCNC-prev-studies}).

For heavier sterile neutrinos, no reliable computations of the
sterile-neutrino impact on BBN were made up to the present time, and the rough
estimate $\tau_s \lesssim 0.1\sec$ is usually used instead (for more details,
see \cite{Dolgov:00b}). This latter bound gives no substantial restriction on
such heavy sterile neutrinos, since from Fig.~\ref{fig:lifetime_result} it can
be inferred that the horizontal line $\tau_s=0.1\sec$ is generally
\emph{higher} than the bounds depicted there.

\begin{figure}[ht]
  \centerline{
  \includegraphics[width=.5\textwidth]{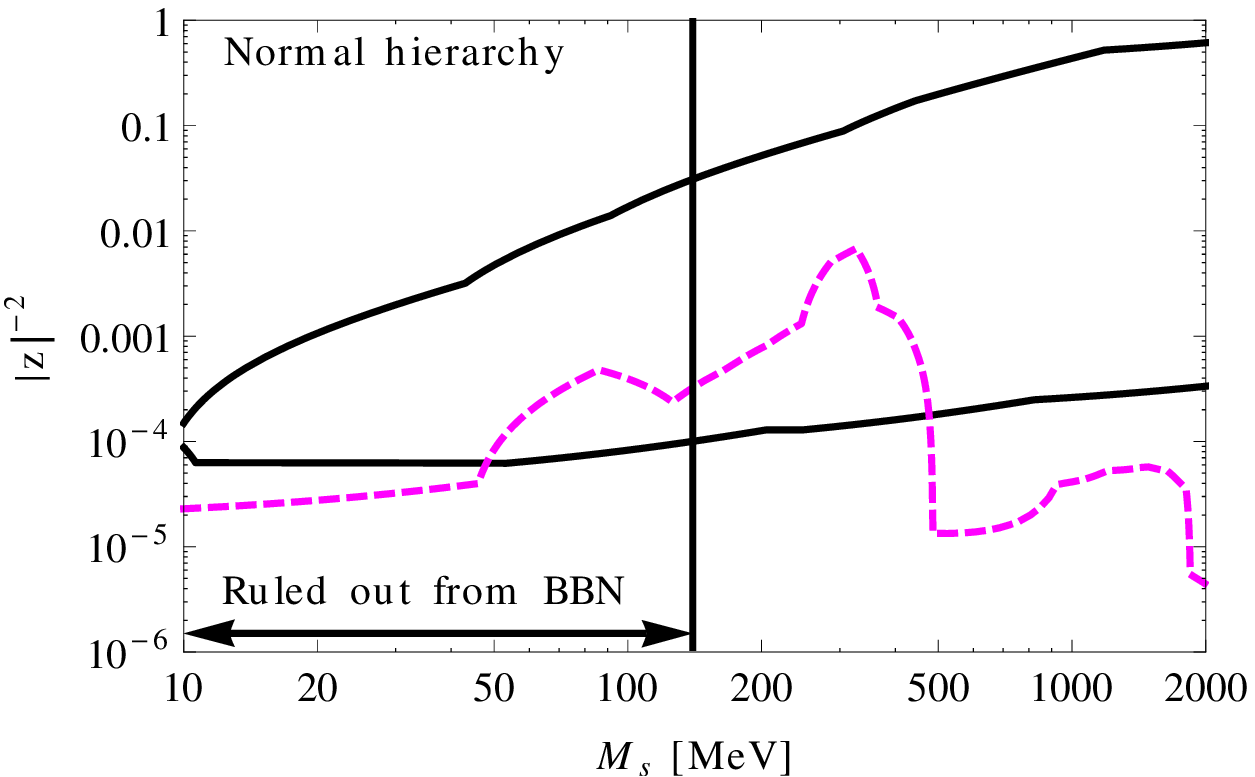}%
  \includegraphics[width=.5\textwidth]{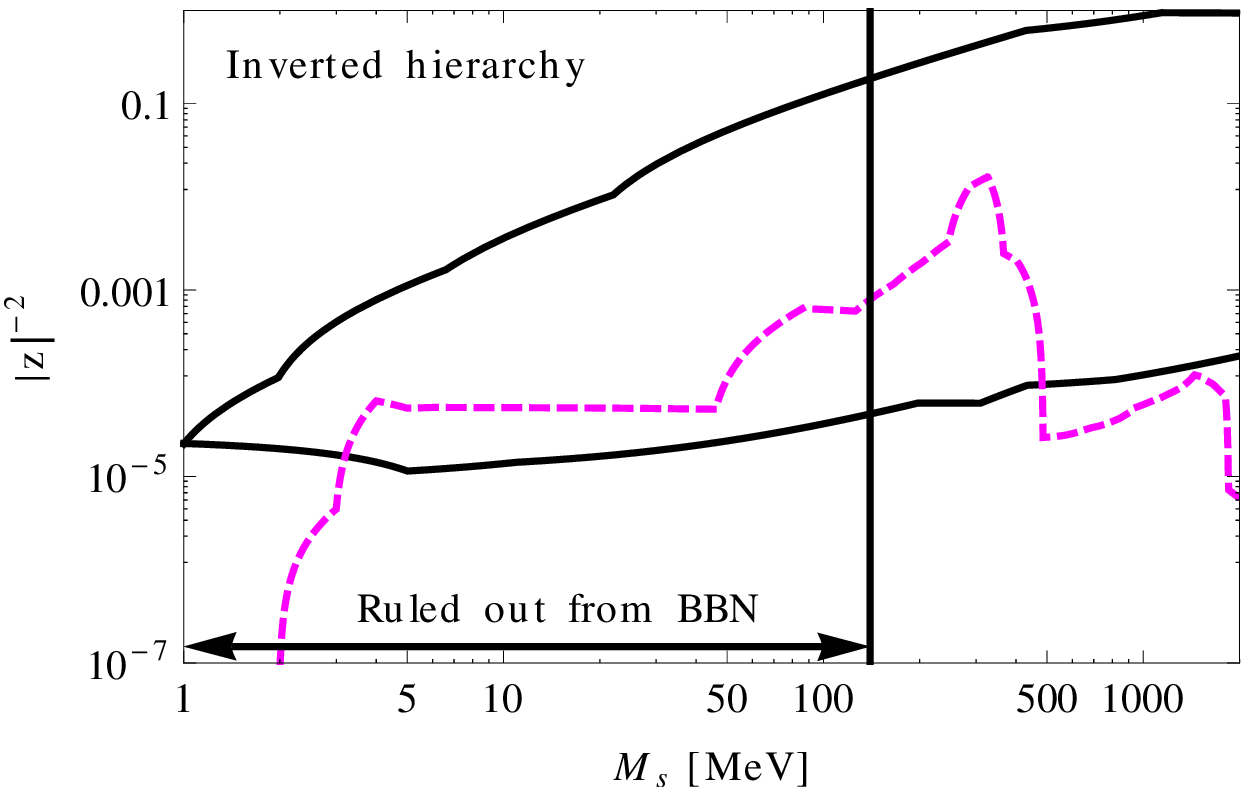}%
}%
\caption{The region of successful baryogenesis in the \numsm compared with the
  experimental upper bounds on the parameter $|z|$. The vales of $M_\m$ and
  $|z|$, lying inside the black solid lines lead to the production of the
  observable baryon asymmetry (from~\protect\cite{Canetti:10a}). The magenta
  dashed line marks is the \emph{lower} bound on $|z|^{-2}$ (parameter, called
  $\epsilon$ in~\protect\cite{Shaposhnikov:08a,Canetti:10a}) such that for
  smaller values at least one of the mixing angles $\vartheta_\alpha^2$ is in
  contradiction with direct experimental searches (for the best-fit values of
  the PMNS mixing angles and masses). The value of $|z|$ corresponding to the
  bound is what we refer to as $z_\text{max}$ in
  Sec.~\protect\ref{sec:minimal-mixing}. The region to the left of $M_s =
  140\mev$ is ruled out from comparison with primordial nucleosynthesis bounds
  (Fig.~\protect\ref{fig:lifetime-BBN}).}
  \label{fig:epsilon_bounds}
\end{figure}

Apart from their influence on BBN, the decays of sterile neutrinos may produce
additional entropy and energy of
plasma~\cite{Fuller:2011qy,Ruchayskiy:2012si}. Their out-of-equilibrium
behaviour may lead as well to the successful baryogenesis
scenario~\cite{Akhmedov:98,Asaka:05b,Canetti:10a}; the generation of large
lepton asymmetry at temperatures below the sphaleron
freeze-out~\cite{Shaposhnikov:08a}. In Fig.~\ref{fig:epsilon_bounds} we
superimpose the bounds on $|z|$, coming from the direct experimental searches
on the region of parameters ($|z|$, $M_\m$) in which the successful
baryogenesis is possible (the region inside the black contours marked ``BAU''
based on the Ref.~\cite{Canetti:10a}).  Additionally, the lepton asymmetry may
be generated at $T\lesssim \mathrm{few}\gev$~\cite{Shaposhnikov:08a} and can
affect production of dark matter sterile
neutrinos~\cite{Shi:98,Laine:08a}. Sterile neutrinos with the mass of about
200~MeV and mixing angles $\sim 10^{-7} - 10^{-8}$ can affect the physics of
supernova explosions~\cite{Fuller:08}.  Finally, our constraints may affect
the low-reheating temperature cosmological scenario (see
e.g.~\cite{Gelmini:04,Gelmini:08}).

\subsubsection*{Acknowledgments.} We would like to thank T.~Asaka,
F.~Bezrukov, A.~Boyarsky, S.~Eijima, D.~Gorbunov, H.~Ishida, S.~Pascoli,
T.~Schwetz-Mangold, D.~Semikoz, M.~Shaposhnikov and J.~Valle for discussion
and useful comments. A.I. is also grateful to S.~Vilchynskiy, Scientific and
Educational Centre of the Bogolyubov Institute for Theoretical Physics in
Kiev, Ukraine\footnote{\url{http://sec.bitp.kiev.ua}} and to Ukrainian Virtual
Roentgen and Gamma-Ray Observatory
VIRGO.UA.\footnote{\url{http://virgo.org.ua}} The work of A.I. was supported
in part from the Swiss-Ukrainian cooperation project (SCOPES)
No.~IZ73Z0\_128040 of Swiss National Science Foundation.

\let\jnlstyle=\rm\def\jref#1{{\jnlstyle#1}}\def\aj{\jref{AJ}}
  \def\araa{\jref{ARA\&A}} \def\apj{\jref{ApJ}\ } \def\apjl{\jref{ApJ}\ }
  \def\apjs{\jref{ApJS}} \def\ao{\jref{Appl.~Opt.}} \def\apss{\jref{Ap\&SS}}
  \def\aap{\jref{A\&A}} \def\aapr{\jref{A\&A~Rev.}} \def\aaps{\jref{A\&AS}}
  \def\azh{\jref{AZh}} \def\baas{\jref{BAAS}} \def\jrasc{\jref{JRASC}}
  \def\memras{\jref{MmRAS}} \def\mnras{\jref{MNRAS}\ }
  \def\pra{\jref{Phys.~Rev.~A}\ } \def\prb{\jref{Phys.~Rev.~B}\ }
  \def\prc{\jref{Phys.~Rev.~C}\ } \def\prd{\jref{Phys.~Rev.~D}\ }
  \def\pre{\jref{Phys.~Rev.~E}} \def\prl{\jref{Phys.~Rev.~Lett.}}
  \def\pasp{\jref{PASP}} \def\pasj{\jref{PASJ}} \def\qjras{\jref{QJRAS}}
  \def\skytel{\jref{S\&T}} \def\solphys{\jref{Sol.~Phys.}}
  \def\sovast{\jref{Soviet~Ast.}} \def\ssr{\jref{Space~Sci.~Rev.}}
  \def\zap{\jref{ZAp}} \def\nat{\jref{Nature}\ } \def\iaucirc{\jref{IAU~Circ.}}
  \def\aplett{\jref{Astrophys.~Lett.}}
  \def\apspr{\jref{Astrophys.~Space~Phys.~Res.}}
  \def\bain{\jref{Bull.~Astron.~Inst.~Netherlands}}
  \def\fcp{\jref{Fund.~Cosmic~Phys.}} \def\gca{\jref{Geochim.~Cosmochim.~Acta}}
  \def\grl{\jref{Geophys.~Res.~Lett.}} \def\jcp{\jref{J.~Chem.~Phys.}}
  \def\jgr{\jref{J.~Geophys.~Res.}}
  \def\jqsrt{\jref{J.~Quant.~Spec.~Radiat.~Transf.}}
  \def\memsai{\jref{Mem.~Soc.~Astron.~Italiana}}
  \def\nphysa{\jref{Nucl.~Phys.~A}} \def\physrep{\jref{Phys.~Rep.}}
  \def\physscr{\jref{Phys.~Scr}} \def\planss{\jref{Planet.~Space~Sci.}}
  \def\procspie{\jref{Proc.~SPIE}} \let\astap=\aap \let\apjlett=\apjl
  \let\apjsupp=\apjs \let\applopt=\ao
\providecommand{\href}[2]{#2}\begingroup\raggedright\endgroup

\appendix

\section{Sterile neutrino lifetime}
\label{app:lifetime}
For any phenomenologically interesting mass of the sterile neutrino it can
decay to three neutrinos $N\rightarrow \nu \bar \nu \nu$ and it results in
contribution
\begin{equation}
B^e_{\nu\nu\nu}=B^\mu_{\nu\nu\nu} = B^\tau_{\nu\nu\nu} =1 
\end{equation}
to the total sum in Eq.~(\ref{eq:lifetime_expression}) over decay products
$X$.

If $M_\m > 2 M_e \approx 1.0\MeV$, then a new decay channel appears: $N
\rightarrow e^+ e^- \nu$. It corresponds to
\begin{align*}
 B^\alpha_{ee\nu} &= (\frac{1}{4} \pm \sin^2 \theta_W + 2 \sin^4 \theta_W) \Biggl[ \left(1- 14 S_e^2 - 2 S_e^4 - 12 S_e^6  \right)\sqrt{1 - 4 S_e^2 } \\
&+ 12  S_e^4 \left( S_e^4 - 1 \right) \ln \biggl( \frac{1 - 3 S_e^2 - (1-S_e^2) \sqrt{1-4S_e^2}}{S_e^2 (1+ \sqrt{1-4S_e^2})} \biggr)  \Biggr]\\
&+ 2 \sin^2 \theta_W (2\sin^2 \theta_W \pm 1) \Biggl[ S_e^2 (2+10 S_e^2 - 12 S_e^4)\sqrt{1-4S_e^2}\\
&+ 6S_e^4 (1-2S_e^2+2S_e^4) \ln \biggl( \frac{1 - 3 S_e^2 - (1-S_e^2) \sqrt{1-4S_e^2}}{S_e^2 (1+ \sqrt{1-4S_e^2})} \biggr) \Biggr]\;,
\end{align*}
where $S_X=M_X/M_\m$, $\theta_W$ is the Weinberg angle, and sign ''+'' corresponds to $\alpha=e$ while ''-'' in the other case. For $S_\mu < \frac{1}{2}$ we should add up to a sum terms $B^\alpha_{\mu\mu\nu}$ that can be obtained from $B^\alpha_{ee\nu}$ if one changes $S_e\rightarrow S_\mu$ and if sign ''+'' corresponds to $\alpha=\mu$, ''-'' - to $\alpha=e,\tau$.

 Other $B$-coefficients are
\begin{equation}
 B^{e,\mu}_{e\mu\nu} = 1 - 8 S_\mu^2 + 8 S_\mu^6 - S_\mu^8 - 24 S_\mu^4 \ln S_\mu,~~B^\tau_{e\mu}=0,~~(S_\mu+S_e < 1)\;,
\end{equation}
\begin{equation}
B^\alpha_{\pi\nu}= 6 \pi^2 \frac{f_\pi^2}{M_\m^2} (1 - S_\pi^2)^2,~~(S_\pi <1,~M_\pi\approx 140\MeV,~f_\pi\approx 130 \MeV)\;,
\end{equation}
\begin{equation}
B^\alpha_{\eta\nu}= 6 \pi^2 \frac{f_\eta^2}{M_\m^2} (1 - S_\eta^2)^2,~~(S_\eta <1,~M_\eta\approx 550\MeV,~f_\eta\approx 155\MeV)\;,
\end{equation}
\begin{equation}
 B^e_{\pi e} = 12 \pi^2 \frac{f_\pi^2}{M_\m^2} V^2_{ud} \biggl( (1-S_e^2)^2 - S_\pi^2 (1+S_e^2) \biggr) \sqrt{(1-(S_\pi-S_e)^2 )(1-(S_\pi+S_e)^2)},~~(S_e+S_\pi<1)\;,
\end{equation}
\begin{equation}
 B^\mu_{\pi \mu} = 12 \pi^2 \frac{f_\pi^2}{M_\m^2} V^2_{ud} \biggl( (1-S_\mu^2)^2 - S_\pi^2 (1+S_\mu^2) \biggr) \sqrt{(1-(S_\pi-S_\mu)^2 )(1-(S_\pi+S_\mu)^2)},~~(S_\mu+S_\pi<1)\;.
\end{equation}
In the last expression quantity $V_{ud} \approx 0.97$ is the element of the
Cabibbo-Kobayashi-Maskawa (CKM) quark matrix.  Expression for $B^e_{Ke}$, that
is non-zero for mass range $S_e+S_K<1$ ($M_K \approx 495\MeV$), can be derived
from $B^e_{\pi e}$ by simultaneous change $S_\pi\rightarrow S_K,
V_{ud}\rightarrow V_{us} \approx 0.23,~f_\pi \rightarrow f_K \approx 160
\MeV$. Moreover, if we change in the resulting expression $S_e$ by $S_\mu$, we
get $B^\mu_{K\mu}$ (for $S_\mu+S_K<1$). What concerns B-coefficients with
$\alpha=\tau$ and $X=(\pi e,~\pi\mu,~Ke,~K\mu)$, they are all zero. For masses
of sterile neutrino, that are larger than the mass of the $\rho$-meson $M_\rho
\approx 780 \MeV$, other $B_X$ appear. We do not list list them in the present
paper, however they can be read from~\cite{Gorbunov:07a}.

\section{PMNS parametrization}
\label{sec:pmns}

In Section~\ref{sec:solution-see-saw} we use the standard definition of the PMNS matrix
(cf. e.g.~\cite{Strumia:06}):
\begin{equation}
\nonumber
V= \left(\begin{array}{ccc} 1&0&0\\ 0&c_{23}&s_{23}\\ 0&-s_{23}&c_{23} \end{array}\right)
\left(\begin{array}{ccc} c_{13}&0&s_{13}\\ 0&e^{i\phi}&0 \\ -s_{13}&0&c_{13} \end{array}\right)
\left(\begin{array}{ccc} c_{12}&s_{12}&0\\ -s_{12}&c_{12}&0 \\ 0&0&1 \end{array}\right) 
\end{equation}
\begin{equation}
 = \left(\begin{array}{ccc} c_{12}c_{13}& c_{13}s_{12} &s_{13}\\
-c_{23}s_{12}e^{i\phi}-c_{12}s_{13}s_{23}& c_{12}c_{23}e^{i\phi} - s_{12}s_{13}s_{23}& c_{13}s_{23}\\
 s_{23}s_{12}e^{i\phi} -c_{12}c_{23}s_{13}& -c_{12}s_{23}e^{i\phi} - c_{23}s_{12}s_{13}& c_{13}c_{23} 
\end{array}\right) \;.
\end{equation}
where $c_{ij}=\cos \theta_{ij}$, and
$s_{ij}=\sin \theta_{ij}$.
The active neutrino mixing matrix is given by expression~(\ref{eq:22}).

\section{Ratio of sterile neutrino mixing angles for $|z|\sim 1$}
\label{sec:z-1corrections}

As the expressions~(\ref{eq:theta_alpha-NH})
and~(\ref{eq:mixing_angles_inverted}) show, the mixing angles have two terms:
one is proportional to $|z|^2$ and another to $|z|^{-2}$ (recall that $|z|\ge
1$).  It was shown in Sec.~\ref{sec:inverted-hierarchy} that for inverted
hierarchy the $|z|^2$-term can be zero for $\vartheta_\mu^2$ and
$\vartheta_\tau^2$, while the $|z|^{-2}$ term in general stays finite.

For a given value of $|z|$, the $|z|^{-2}$-term is bounded from
above. According to (\ref{eq:theta_alpha-NH})
and~(\ref{eq:mixing_angles_inverted}), its maximum is realized simultaneously
with the maximal value of
\begin{equation}
L^{NH}_\alpha=| V_{\alpha 3} + i e^{i\xi}\sqrt{\frac{m_2}{m_3}}V_{\alpha 2}|^2
\label{eq:L_alpha-NH}
\end{equation}
in the normal hierarchy, and
\begin{equation}
L^{IH}_\alpha= |V_{\alpha 1} + ie^{i(\xi-\zeta)}\sqrt{\frac{m_2}{m_1}} V_{\alpha 2}|^2
\label{eq:L_alpha-IH}
\end{equation}
in the inverted hierarchy.

Analysis, similar to that of the
Sections~\ref{sec:normal-hierarchy}--\ref{sec:inverted-hierarchy} shows that
\begin{equation}
L^{NH}_e \leq 0.2,~~L^{NH}_\mu \leq 1.1,~~L^{NH}_\tau \leq 1.1,~~L^{IH}_e\leq 1.96,~~L^{IH}_\mu\leq 1.3,~~L^{IH}_\tau\leq 1.3 \;.
\label{eq:L_alpha-bounds}
\end{equation}
These bounds allow to estimate the contribution of the $|z|^{-2}$-terms to the
whole sum of the squared mixing angles
\begin{equation}
\sum \limits_\alpha \vartheta_\alpha^2 = \frac{m_1+m_2+m_3}{4 M_\m} \left(|z|² + \frac1{|z|²} \right)\;.
\label{eq:sum-angles-accurate}
\end{equation}
The ratio of~(\ref{eq:L_alpha-NH})--(\ref{eq:L_alpha-IH})
to~(\ref{eq:sum-angles-accurate}) gives
\begin{equation}
R^{NH}_\alpha =\frac{L^{NH}_\alpha}{ (1+\frac{m_2}{m_3}) }\frac1{|z|^4+1},~~R^{IH}_\alpha =\frac{L^{IH}_\alpha}{ (1+\frac{m_2}{m_1}) }\frac1{|z|^4+1} \;.
\end{equation}
For $z\sim 1$ it can become of order unity. 
However, we restrict ourselves to
the sufficiently large values $z\gtrsim 10$, that are consistent with the
upper bound, indicated by the experiments (see Fig.\ref{fig:epsilon_bounds})
\begin{equation}
R^{NH}_e \lesssim 2\times 10^{-5},~~R^{NH}_{\mu,\tau} \lesssim
10^{-4},~~R^{IH}_e\lesssim 10^{-4},~~R^{IH}_{\mu,\tau}\lesssim 5\times10^{-5} \;.
\end{equation}
Comparison these results with the {\it lower} bounds
(Table~\ref{tab:mixing-ratio-bounds}) we see that $z^{-1}$ terms are
unimportant for the for all mixing angles in NH and $\vartheta_e$ in IH. What
concerns the remaining angles $\vartheta_\mu$ and $\vartheta_\tau$ in IH, they
can be substantially modified by account of $z^{-1}$-terms, but anyway each of
them can become small enough, compared to the other angles, as explained in
next section. As a corollary, analysis and results of
Secs. \ref{sec:normal-hierarchy},\ref{sec:inverted-hierarchy} do not change
significantly for large enough values of $z$.


\begin{thebibliography}{10}

\bibitem{Strumia:06}
A.~Strumia and F.~Vissani, {\it Neutrino masses and mixings and.},
  \href{http://arXiv.org/abs/hep-ph/0606054}{{\tt hep-ph/0606054}}.

\bibitem{Schwetz:08a}
T.~Schwetz, M.~Tortola and J.~W.~F. Valle, {\it {Three-flavour neutrino
  oscillation update}},  {\em New J. Phys.} {\bf 10} (2008) 113011
  [\href{http://arXiv.org/abs/0808.2016}{{\tt 0808.2016}}].

\bibitem{Schwetz:2011zk}
T.~Schwetz, M.~Tortola and J.~Valle, {\it {Where we are on $\theta_{13}$:
  addendum to 'Global neutrino data and recent reactor fluxes: status of
  three-flavour oscillation parameters'}},  {\em New J.Phys.} {\bf 13} (2011)
  109401 [\href{http://arXiv.org/abs/1108.1376}{{\tt 1108.1376}}].

\bibitem{Fogli:2011qn}
G.~L. Fogli, E.~Lisi, A.~Marrone, A.~Palazzo and A.~M. Rotunno, {\it {Evidence
  of theta(13)>0 from global neutrino data analysis}},  {\em Phys.Rev.} {\bf
  D84} (2011) 053007 [\href{http://arXiv.org/abs/1106.6028}{{\tt 1106.6028}}].

\bibitem{PDG:11}
{\bf Particle Data Group} Collaboration, K.~Nakamura {\em et.~al.}, {\it
  {Review of particle physics}},  {\em J.Phys.G} {\bf G37} (2010) 075021.

\bibitem{WMAP7}
E.~{Komatsu}, K.~M. {Smith}, J.~{Dunkley}, C.~L. {Bennett}, B.~{Gold},
  G.~{Hinshaw}, N.~{Jarosik}, D.~{Larson}, M.~R. {Nolta}, L.~{Page}, D.~N.
  {Spergel}, M.~{Halpern}, R.~S. {Hill}, A.~{Kogut}, M.~{Limon}, S.~S. {Meyer},
  N.~{Odegard}, G.~S. {Tucker}, J.~L. {Weiland}, E.~{Wollack} and E.~L.
  {Wright}, {\it {Seven-year Wilkinson Microwave Anisotropy Probe (WMAP)
  Observations: Cosmological Interpretation}},  {\em \apjs} {\bf 192} (Feb.,
  2011) 18--+ [\href{http://arXiv.org/abs/1001.4538}{{\tt 1001.4538}}].

\bibitem{Minkowski:77}
P.~Minkowski, {\it mu $\to$ e gamma at a rate of one out of 1-billion muon
  decays?},  {\em Phys. Lett.} {\bf B67} (1977) 421.

\bibitem{Ramond:79}
P.~Ramond, {\it The family group in grand unified theories},
  \href{http://arXiv.org/abs/hep-ph/9809459}{{\tt hep-ph/9809459}}.

\bibitem{Mohapatra:79}
R.~N. Mohapatra and G.~Senjanovic, {\it Neutrino mass and spontaneous parity
  nonconservation},  {\em Phys. Rev. Lett.} {\bf 44} (1980) 912.

\bibitem{Yanagida:80}
T.~Yanagida, {\it Horizontal gauge symmetry and masses of neutrinos},  {\em
  Prog. Theor. Phys.} {\bf 64} (1980) 1103.

\bibitem{Atre:09}
A.~Atre, T.~Han, S.~Pascoli and B.~Zhang, {\it {The Search for Heavy Majorana
  Neutrinos}},  {\em JHEP} {\bf 05} (2009) 030
  [\href{http://arXiv.org/abs/0901.3589}{{\tt 0901.3589}}].

\bibitem{Gorbunov:07a}
D.~Gorbunov and M.~Shaposhnikov, {\it How to find neutral leptons of the
  numsm?},  {\em JHEP} {\bf 10} (2007) 015
  [\href{http://arXiv.org/abs/arXiv:0705.1729 [hep-ph]}{{\tt arXiv:0705.1729
  [hep-ph]}}].

\bibitem{Shaposhnikov:08a}
M.~{Shaposhnikov}, {\it {The nuMSM, leptonic asymmetries, and properties of
  singlet fermions}},  {\em JHEP} {\bf 08} (2008) 008
  [\href{http://arXiv.org/abs/0804.4542}{{\tt 0804.4542}}].

\bibitem{Dolgov:00a}
A.~D. Dolgov, S.~H. Hansen, G.~Raffelt and D.~V. Semikoz, {\it {Cosmological
  and astrophysical bounds on a heavy sterile neutrino and the KARMEN
  anomaly}},  {\em Nucl. Phys.} {\bf B580} (2000) 331--351
  [\href{http://arXiv.org/abs/hep-ph/0002223}{{\tt hep-ph/0002223}}].

\bibitem{Dolgov:00b}
A.~D. Dolgov, S.~H. Hansen, G.~Raffelt and D.~V. Semikoz, {\it {Heavy sterile
  neutrinos: Bounds from big-bang nucleosynthesis and SN 1987A}},  {\em Nucl.
  Phys.} {\bf B590} (2000) 562--574
  [\href{http://arXiv.org/abs/hep-ph/0008138}{{\tt hep-ph/0008138}}].

\bibitem{Boyarsky:09a}
A.~Boyarsky, O.~Ruchayskiy and M.~Shaposhnikov, {\it {The role of sterile
  neutrinos in cosmology and astrophysics}},  {\em Ann. Rev. Nucl. Part. Sci.}
  {\bf 59} (2009) 191 [\href{http://arXiv.org/abs/0901.0011}{{\tt 0901.0011}}].

\bibitem{Asaka:2011pb}
T.~Asaka, S.~Eijima and H.~Ishida, {\it {Mixing of Active and Sterile
  Neutrinos}},  {\em JHEP} {\bf 1104} (2011) 011
  [\href{http://arXiv.org/abs/1101.1382}{{\tt 1101.1382}}].

\bibitem{An:2012eh}
{\bf DAYA-BAY Collaboration} Collaboration, F.~An {\em et.~al.}, {\it
  {Observation of electron-antineutrino disappearance at Daya Bay}},
  \href{http://arXiv.org/abs/1203.1669}{{\tt 1203.1669}}.

\bibitem{Ahn:2012nd}
{\bf RENO collaboration} Collaboration, J.~Ahn {\em et.~al.}, {\it {Observation
  of Reactor Electron Antineutrino Disappearance in the RENO Experiment}},
  \href{http://arXiv.org/abs/1204.0626}{{\tt 1204.0626}}.

\bibitem{Adamson:2011qu}
{\bf MINOS} Collaboration, P.~Adamson {\em et.~al.}, {\it {Improved search for
  muon-neutrino to electron-neutrino oscillations in MINOS}},  {\em
  Phys.Rev.Lett.} {\bf 107} (2011) 181802
  [\href{http://arXiv.org/abs/1108.0015}{{\tt 1108.0015}}].

\bibitem{Abe:2011sj}
{\bf T2K} Collaboration, K.~Abe {\em et.~al.}, {\it {Indication of Electron
  Neutrino Appearance from an Accelerator-produced Off-axis Muon Neutrino
  Beam}},  {\em Phys.Rev.Lett.} {\bf 107} (2011) 041801
  [\href{http://arXiv.org/abs/1106.2822}{{\tt 1106.2822}}].

\bibitem{Ruchayskiy:2012si}
O.~Ruchayskiy and A.~Ivashko, {\it {Restrictions on the lifetime of sterile
  neutrinos from primordial nucleosynthesis}},
  \href{http://arXiv.org/abs/1202.2841}{{\tt 1202.2841}}.

\bibitem{Gorkavenko:2009vd}
V.~Gorkavenko and S.~Vilchynskiy, {\it {Some constraints on the Yukawa
  parameters in the neutrino modification of the Standard Model (nuMSM) and
  CP-violation}},  {\em Eur.Phys.J.} {\bf C70} (2010) 1091--1098
  [\href{http://arXiv.org/abs/0907.4484}{{\tt 0907.4484}}].

\bibitem{Seesaw25}
World Scientific, {\em International Conference on the Seesaw Mechanism},
  (Singapore), World Scientific, 2005.

\bibitem{Schechter:1980gr}
J.~Schechter and J.~Valle, {\it {Neutrino Masses in SU(2) x U(1) Theories}},
  {\em Phys.Rev.} {\bf D22} (1980) 2227.

\bibitem{Schechter:1981cv}
J.~Schechter and J.~Valle, {\it {Neutrino Decay and Spontaneous Violation of
  Lepton Number}},  {\em Phys.Rev.} {\bf D25} (1982) 774.

\bibitem{Rodejohann:2011vc}
W.~Rodejohann and J.~Valle, {\it {Symmetrical Parametrizations of the Lepton
  Mixing Matrix}},  {\em Phys.Rev.} {\bf D84} (2011) 073011
  [\href{http://arXiv.org/abs/1108.3484}{{\tt 1108.3484}}].

\bibitem{Asaka:05a}
T.~Asaka, S.~Blanchet and M.~Shaposhnikov, {\it The numsm, dark matter and
  neutrino masses},  {\em Phys. Lett.} {\bf B631} (2005) 151--156
  [\href{http://arXiv.org/abs/hep-ph/0503065}{{\tt hep-ph/0503065}}].

\bibitem{Asaka:05b}
T.~{Asaka} and M.~{Shaposhnikov}, {\it {The nuMSM, dark matter and baryon
  asymmetry of the universe [rapid communication]}},  {\em Phys. Lett. B} {\bf
  620} (July, 2005) 17--26
  [\href{http://arXiv.org/abs/arXiv:hep-ph/0505013}{{\tt
  arXiv:hep-ph/0505013}}].

\bibitem{Boyarsky:06a}
A.~Boyarsky, A.~Neronov, O.~Ruchayskiy and M.~Shaposhnikov, {\it The masses of
  active neutrinos in the {nuMSM} from {X}-ray astronomy},  {\em JETP Letters}
  (2006) 133--135 [\href{http://arXiv.org/abs/hep-ph/0601098}{{\tt
  hep-ph/0601098}}].

\bibitem{Canetti:10a}
L.~Canetti and M.~Shaposhnikov, {\it {Baryon Asymmetry of the Universe in the
  NuMSM}},  {\em JCAP} {\bf 1009} (2010) 001
  [\href{http://arXiv.org/abs/1006.0133}{{\tt 1006.0133}}].

\bibitem{Shaposhnikov:06b}
M.~Shaposhnikov, {\it A possible symmetry of the numsm},  {\em Nucl. Phys.}
  {\bf B763} (2007) 49--59 [\href{http://arXiv.org/abs/hep-ph/0605047}{{\tt
  hep-ph/0605047}}].

\bibitem{Casas:01}
J.~A. Casas and A.~Ibarra, {\it {Oscillating neutrinos and mu --> e, gamma}},
  {\em Nucl. Phys.} {\bf B618} (2001) 171--204
  [\href{http://arXiv.org/abs/hep-ph/0103065}{{\tt hep-ph/0103065}}].

\bibitem{Bernardi:1985ny}
G.~Bernardi {\em et.~al.}, {\it {SEARCH FOR NEUTRINO DECAY}},  {\em Phys.
  Lett.} {\bf B166} (1986) 479.

\bibitem{Bernardi:1987ek}
G.~Bernardi {\em et.~al.}, {\it Further limits on heavy neutrino couplings},
  {\em Phys. Lett.} {\bf B203} (1988) 332.

\bibitem{Britton:1992pg}
D.~Britton, S.~Ahmad, D.~Bryman, R.~Burnbam, E.~Clifford {\em et.~al.}, {\it
  {Measurement of the $\pi^+ \to e^+ \nu$ branching ratio}},  {\em
  Phys.Rev.Lett.} {\bf 68} (1992) 3000--3003.

\bibitem{Britton:1992xv}
D.~Britton, S.~Ahmad, D.~Bryman, R.~Burnham, E.~Clifford {\em et.~al.}, {\it
  {Improved search for massive neutrinos in $\pi^+ \to e^+ \nu$ decay}},  {\em
  Phys.Rev.} {\bf D46} (1992) 885--887.

\bibitem{Vaitaitis:1999wq}
{\bf NuTeV} Collaboration, A.~Vaitaitis {\em et.~al.}, {\it {Search for neutral
  heavy leptons in a high-energy neutrino beam}},  {\em Phys. Rev. Lett.} {\bf
  83} (1999) 4943--4946 [\href{http://arXiv.org/abs/hep-ex/9908011}{{\tt
  hep-ex/9908011}}].

\bibitem{CHARM:1985}
{\bf CHARM} Collaboration, F.~Bergsma {\em et.~al.}, {\it {A SEARCH FOR DECAYS
  OF HEAVY NEUTRINOS IN THE MASS RANGE 0.5-GeV TO 2.8-GeV}},  {\em Phys.Lett.}
  {\bf B166} (1986) 473.

\bibitem{Yamazaki:1984sj}
T.~Yamazaki {\em et.~al.}, {\it Search for heavy neutrinos in kaon decay}, . IN
  *LEIPZIG 1984, PROCEEDINGS, HIGH ENERGY PHYSICS, VOL. 1*, 262.

\bibitem{Hayano:1982wu}
R.~Hayano, T.~Taniguchi, T.~Yamanaka, T.~Tanimori, R.~Enomoto {\em et.~al.},
  {\it {HEAVY NEUTRINO SEARCH USING K(mu2) DECAY}},  {\em Phys.Rev.Lett.} {\bf
  49} (1982) 1305.

\bibitem{Bryman:1996xd}
D.~Bryman and T.~Numao, {\it {Search for massive neutrinos in $\pi+ \to \mu +
  \nu$ decay}},  {\em Phys.Rev.} {\bf D53} (1996) 558--559.

\bibitem{Abela:1981nf}
R.~Abela, M.~Daum, G.~Eaton, R.~Frosch, B.~Jost {\em et.~al.}, {\it {SEARCH FOR
  AN ADMIXTURE OF HEAVY NEUTRINO IN PION DECAY}},  {\em Phys.Lett.} {\bf B105}
  (1981) 263--266.

\bibitem{Daum:1987bg}
M.~Daum, B.~Jost, R.~Marshall, R.~Minehart, W.~Stephens {\em et.~al.}, {\it
  {SEARCH FOR ADMIXTURES OF MASSIVE NEUTRINOS IN THE DECAY pi+ ---> mu+
  neutrino}},  {\em Phys.Rev.} {\bf D36} (1987) 2624.

\bibitem{Aoki:2011vma}
{\bf PIENU Collaboration} Collaboration, M.~Aoki {\em et.~al.}, {\it {Search
  for Massive Neutrinos in the Decay $\pi \to e \nu$}},  {\em Phys.Rev.} {\bf
  D84} (2011) 052002 [\href{http://arXiv.org/abs/1106.4055}{{\tt 1106.4055}}].

\bibitem{Abreu:1996pa}
{\bf DELPHI} Collaboration, P.~Abreu {\em et.~al.}, {\it {Search for neutral
  heavy leptons produced in Z decays}},  {\em Z.Phys.} {\bf C74} (1997) 57--71.

\bibitem{Blennow:2010th}
M.~Blennow, E.~Fernandez-Martinez, J.~Lopez-Pavon and J.~Menendez, {\it
  {Neutrinoless double beta decay in seesaw models}},  {\em JHEP} {\bf 1007}
  (2010) 096 [\href{http://arXiv.org/abs/1005.3240}{{\tt 1005.3240}}].

\bibitem{Bezrukov:2005mx}
F.~Bezrukov, {\it numsm predictions for neutrinoless double beta decay},  {\em
  Phys. Rev.} {\bf D72} (2005) 071303
  [\href{http://arXiv.org/abs/hep-ph/0505247}{{\tt hep-ph/0505247}}].

\bibitem{Shrock:80}
R.~E. Shrock, {\it {New Tests For, and Bounds On, Neutrino Masses and Lepton
  Mixing}},  {\em Phys. Lett.} {\bf B96} (1980) 159.

\bibitem{Shrock:1980ct}
R.~E. Shrock, {\it {General Theory of Weak Leptonic and Semileptonic Decays. 1.
  Leptonic Pseudoscalar Meson Decays, with Associated Tests For, and Bounds on,
  Neutrino Masses and Lepton Mixing}},  {\em Phys.Rev.} {\bf D24} (1981) 1232.

\bibitem{Shrock:1981wq}
R.~E. Shrock, {\it {General Theory of Weak Processes Involving Neutrinos. 2.
  Pure Leptonic Decays}},  {\em Phys.Rev.} {\bf D24} (1981) 1275.

\bibitem{Shrock:1981cq}
R.~E. Shrock, {\it {PURE LEPTONIC DECAYS WITH MASSIVE NEUTRINOS AND ARBITRARY
  LORENTZ STRUCTURE}},  {\em Phys.Lett.} {\bf B112} (1982) 382.

\bibitem{Kusenko:2004qc}
A.~Kusenko, S.~Pascoli and D.~Semikoz, {\it {New bounds on MeV sterile
  neutrinos based on the accelerator and Super-Kamiokande results}},  {\em
  JHEP} {\bf 0511} (2005) 028 [\href{http://arXiv.org/abs/hep-ph/0405198}{{\tt
  hep-ph/0405198}}].

\bibitem{Astier:2001ck}
{\bf NOMAD} Collaboration, P.~Astier {\em et.~al.}, {\it Search for heavy
  neutrinos mixing with tau neutrinos},  {\em Phys. Lett.} {\bf B506} (2001)
  27--38 [\href{http://arXiv.org/abs/hep-ex/0101041}{{\tt hep-ex/0101041}}].

\bibitem{Levy}
J.-M. Levy, {\it Doctoral thesis, university of paris (1986)}, .

\bibitem{Akiri:2011dv}
{\bf LBNE Collaboration} Collaboration, T.~Akiri {\em et.~al.}, {\it {The 2010
  Interim Report of the Long-Baseline Neutrino Experiment Collaboration Physics
  Working Groups}},  \href{http://arXiv.org/abs/1110.6249}{{\tt 1110.6249}}.

\bibitem{Abazajian:2012ys}
K.~Abazajian, M.~Acero, S.~Agarwalla, A.~Aguilar-Arevalo, C.~Albright {\em
  et.~al.}, {\it {Light Sterile Neutrinos: A White Paper}},
  \href{http://arXiv.org/abs/1204.5379}{{\tt 1204.5379}}.

\bibitem{Shaykhiev:2011zz}
A.~Shaykhiev, Y.~Kudenko and A.~Khotyantsev, {\it {Searches for heavy neutrinos
  in the decays of positively charged kaons}},  {\em Phys.Atom.Nucl.} {\bf 74}
  (2011) 788--793.

\bibitem{Kusenko:06a}
A.~Kusenko, {\it Sterile neutrinos, dark matter, and the pulsar velocities in
  models with a higgs singlet},  {\em Phys. Rev. Lett.} {\bf 97} (2006) 241301
  [\href{http://arXiv.org/abs/hep-ph/0609081}{{\tt hep-ph/0609081}}].

\bibitem{Shoemaker:2010fg}
I.~M. Shoemaker, K.~Petraki and A.~Kusenko, {\it {Collider signatures of
  sterile neutrinos in models with a gauge-singlet Higgs}},  {\em JHEP} {\bf
  1009} (2010) 060 [\href{http://arXiv.org/abs/1006.5458}{{\tt 1006.5458}}].

\bibitem{Dunkley:2010ge}
J.~Dunkley, R.~Hlozek, J.~Sievers, V.~Acquaviva, P.~Ade {\em et.~al.}, {\it
  {The Atacama Cosmology Telescope: Cosmological Parameters from the 2008 Power
  Spectra}},  {\em Astrophys.J.} {\bf 739} (2011) 52
  [\href{http://arXiv.org/abs/1009.0866}{{\tt 1009.0866}}].

\bibitem{Aver:2011bw}
E.~Aver, K.~A. Olive and E.~D. Skillman, {\it {An MCMC determination of the
  primordial helium abundance}},  \href{http://arXiv.org/abs/1112.3713}{{\tt
  1112.3713}}.

\bibitem{Izotov:2010ca}
Y.~Izotov and T.~Thuan, {\it {The primordial abundance of 4He: evidence for
  non-standard big bang nucleosynthesis}},  {\em Astrophys.J.} {\bf 710} (2010)
  L67--L71 [\href{http://arXiv.org/abs/1001.4440}{{\tt 1001.4440}}].

\bibitem{Fuller:2011qy}
G.~M. Fuller, C.~T. Kishimoto and A.~Kusenko, {\it {Heavy sterile neutrinos,
  entropy and relativistic energy production, and the relic neutrino
  background}},  \href{http://arXiv.org/abs/1110.6479}{{\tt 1110.6479}}.

\bibitem{Akhmedov:98}
E.~K. Akhmedov, V.~A. Rubakov and A.~Y. Smirnov, {\it Baryogenesis via neutrino
  oscillations},  {\em Phys. Rev. Lett.} {\bf 81} (1998) 1359--1362
  [\href{http://arXiv.org/abs/hep-ph/9803255}{{\tt hep-ph/9803255}}].

\bibitem{Shi:98}
X.-d. Shi and G.~M. Fuller, {\it A new dark matter candidate: Non-thermal
  sterile neutrinos},  {\em Phys. Rev. Lett.} {\bf 82} (1999) 2832--2835
  [\href{http://arXiv.org/abs/astro-ph/9810076}{{\tt astro-ph/9810076}}].

\bibitem{Laine:08a}
M.~{Laine} and M.~{Shaposhnikov}, {\it {Sterile neutrino dark matter as a
  consequence of {$\nu$}MSM-induced lepton asymmetry}},  {\em JCAP} {\bf 6}
  (June, 2008) 31--+ [\href{http://arXiv.org/abs/arXiv:0804.4543}{{\tt
  arXiv:0804.4543}}].

\bibitem{Fuller:08}
G.~M. Fuller, A.~Kusenko and K.~Petraki, {\it {Eosphoric sterile neutrinos,
  supernovae, and the galactic positrons}},  {\em Phys. Lett.} {\bf B670}
  (2009) 281--284 [\href{http://arXiv.org/abs/0806.4273}{{\tt 0806.4273}}].

\bibitem{Gelmini:04}
G.~Gelmini, S.~Palomares-Ruiz and S.~Pascoli, {\it Low reheating temperature
  and the visible sterile neutrino},  {\em Phys. Rev. Lett.} {\bf 93} (2004)
  081302 [\href{http://arXiv.org/abs/astro-ph/0403323}{{\tt
  astro-ph/0403323}}].

\bibitem{Gelmini:08}
G.~Gelmini, E.~Osoba, S.~Palomares-Ruiz and S.~Pascoli, {\it {MeV sterile
  neutrinos in low reheating temperature cosmological scenarios}},  {\em JCAP}
  {\bf 0810} (2008) 029 [\href{http://arXiv.org/abs/0803.2735}{{\tt
  0803.2735}}].

\end{thebibliography}
\end{document}